\documentclass[12pt]{article}
\usepackage{natbib}
\usepackage{amsmath, tikz, graphicx, subfigure}
\usepackage{lscape}
\usepackage[labelsep=period,labelfont=bf]{caption}
\usepackage[top=1in, bottom=1in, left=1.25in, right=1.25in]{geometry}
\usetikzlibrary{arrows,positioning}

\newtheorem{example}{Example}
\newtheorem{crit}{Criterion}

\newtheorem{condition}{Condition}
\newtheorem{defn}{Definition}
\newtheorem{prop}{Proposition}

\newtheorem{nono-crit}{Criterion}

\title{On the Individual Surrogate Paradox}

\author{Linquan Ma$^{1,3}$, Yunjian Yin$^{2}$, Lan Liu$^{3}$ and Zhi Geng$^{2}$\\ \\
	$^{1}${\small University of Wisconsin-Madison, Madison, Wisconsin, USA}\\
	$^{2}${\small School of Mathematical Sciences, Peking University, Beijing, China}\\
	$^{3}${\small School of Statistics, University of Minnesota at Twin Cities, Minneapolis, Minnesota, USA}}
\date{}

\begin{document}
\maketitle	

\label{firstpage}

%  put the summary for your paper here

\begin{abstract}
When the primary outcome is difficult to collect, surrogate endpoint is typically used as a substitute. It is possible that for every individual, treatment has a positive effect on surrogate, and surrogate has a positive effect on primary outcome, but for some individuals, treatment has a negative effect on primary outcome. For example, a treatment may be substantially effective in preventing the stroke for everyone, and lowering the risk of stroke is universally beneficial for a longer survival time, however, the treatment may still causes death for some individuals. We define such paradoxical phenomenon as individual surrogate paradox. The individual surrogate paradox is proposed to capture the treatment effect heterogeneity, which is unable to be described by either the surrogate paradox based on average causal effect (ACE) \citep{chen2007criteria} or that based on distributional causal effect (DCE) \citep{ju2010criteria}. We investigate existing surrogate criteria in terms of whether the individual surrogate paradox could manifest. We find that only the strong binary surrogate can avoid such paradox without additional assumptions. Utilizing the sharp bounds, we propose novel criteria to exclude the individual surrogate paradox. Our methods are illustrated in an application to determine the effect of the intensive glycemia on the risk of development or progression of diabetic retinopathy.% using glycated hemoglobin level as a surrogate. 
\end{abstract}

\textbf{keywords:} Biomarkers; Heterogeneity; Individual surrogate paradox; Surrogate endpoints.

\section{Introduction}
\label{s:intro}
To reduce costs and durations of studies and to facilitate collection, a surrogate endpoint or a biomarker is typically used as a substitute for the primary endpoint.  For example, in a cancer trial, the primary outcome is elimination of symptomatic disease or reduction in mortality, and a surrogate endpoint can be chosen as the presence of cancer shown by biopsy during follow up \citep{fleming1996surrogate}. In medicine and social sciences, it is critical to choose a good surrogate endpoint since a bad surrogate would result in misleading or severe consequences \citep{gordon1995cholesterol,moore1997deadly}. For example, while being effective in reducing the risk for arrythmia,  anti-arrythmia drug Tamnbocor directly caused the death of over 50,000 people \citep{moore1997deadly}. 

Various criteria have been proposed to evaluate the use of surrogate endpoint \citep{buyse1998criteria,buyse2000validation,daniels1997meta,molenberghs2001evaluation,freedman1992statistical}. The Prentice criterion \citep{prentice1989surrogate} requires the treatment to be independent with the primary endpoint conditional on surrogate. However, the Prentice criterion might not satisfy the property of causal necessity, that is, the absence of the treatment effect on the surrogate may not indicate the absence of the treatment effect on the primary outcome \citep{frangakis2002principal}. The principal surrogate criterion was proposed by \citet{frangakis2002principal} to satisfy the causal necessity and it is further investigated by \citet{elliott2013discussion,conlon2014surrogacy,gilbert2008evaluating,huang2011comparing,wolfson2010statistical}.  \citet{pearl2010Transportability} demonstrated that the principal surrogacy criterion still cannot  guarantee a surrogate to be a good predictor of the treatment effect on the outcome. \citet{lauritzen2004discussion} proposed a strong surrogate criterion, which is the treatment affects the primary outcome only through the surrogate endpoint. The strong surrogates satisfy causal necessity and are also principal surrogates. % That is, the strong surrogate fully mediates the effect of treatment on the outcome. By definition, A strong surrogate satisfies causal necessity and is also a principal surrogate. gilbert2008evaluating_hudgens,

In recent years, these criteria have been scrutinized from the perspective of whether paradoxical result would manifest. As a minimal requirement,  inference based on a good surrogate endpoint should provide at least the correct sign,  if not the accurate magnitude, of the treatment effect on the primary outcome. \citet{chen2007criteria} pointed out that all of the above criteria suffer from the surrogate paradox based on average causal effect (ACE). That is, for a surrogate that has a positive ACE on the true endpoint, a positive ACE of treatment on the surrogate may still result in a negative ACE of treatment on the primary outcome. Such paradoxical phenomenon happens even in randomized control trials. The authors further proposed criterion to exclude the surrogate paradox based on ACE \citep{chen2007criteria}. % since although the treatment is randomized, the surrogate and primary outcome may still be confounded by some unmeasured confounders. %Assuming monotonicity in the distribution functions of the treatment, surrogate, outcome and the unmeasured confounders, \citet{chen2007criteria} proposed the criterion to exclude the ACE surrogate paradox. 
 \citet{ju2010criteria} demonstrated that a surrogate which can avoid the surrogate paradox based on ACE may still encounter the surrogate paradox in the distributional sense. Based on this, they introduced the surrogate paradox based on distributional causal effect (DCE), where the ACE was replaced with a distributional causal effect, i.e., the difference between the cumulative distribution functions of the two potential outcomes. They also proposed criteria to exclude the surrogate paradox based on DCE. More discussion of the two surrogate paradoxes can be found in \citet{wu2011sufficient} and \citet{vanderweele2013surrogate}. %The absence of surrogate paradox based on ACE and DCE have also been used as criteria to evaluate surrogate endpoints. %Under semiparametric and nonparametric assumptions, they proposed sufficient conditions to exclude the DCE surrogate paradox. This result was further generalized by \citet{wu2011sufficient} and \citet{vanderweele2013surrogate} to the non-strong surrogate case (\textbf{CHECK}). 

% Various criteria have been proposed to exclude the surrogate paradox including but not limited to \citet{chen2007criteria,wu2011sufficient,ju2010criteria,vanderweele2013surrogate}.

%The surrogate paradox raised by \citet{chen2007criteria} was based on the average causal effect (ACE). 
However, the paradoxes based on both ACE and DCE quantify the causal relationship between variables using population measures: average causal effect for the former and difference between distributions of potential outcomes for the latter. Thus, both paradoxes focus on the performance of surrogate endpoint in the entire population rather than  reflecting the individual status. The differences between individuals in underlying pathology, biology or genetics can lead to drastically heterogeneous surrogate performance. For a good surrogate, if every individual gets a beneficial treatment effect on surrogate and a beneficial surrogate effect on the primary endpoint, then it is expected that the no individuals would get a harmful treatment effect on the primary endpoint. More generally, the conclusion of beneficial treatment effect for majority of individuals based on surrogate endpoint should also apply to majority of individuals in terms of primary endpoint. Otherwise, we say the individual surrogate paradox manifests. %, if the surrogate has a positive effect on the primary outcome and the treatment has a positive effect on the surrogate for almost every individual, but the treatment has a negative effect on the primary endpoint for a notable proportion of individuals. In other words, even a surrogate is predictive for the primary endpoint for every individual, it is still possible that the evaluation of the treatment based on the surrogate endpoint has paradoxical conclusion on a subgroup.  
For example, consider a study of Warfarin therapy, where prevention of stroke is the surrogate endpoint and the survival is the primary outcome \citep{borosak2004warfarin}. %stroke is the third leading cause of death in the United States. 
It is found that Warfarin therapy is clinically substantially effective in %the overall prevention of arterial and venous thrombosis, and in 
the primary and secondary prevention of stroke related to non-rheumatic atrial fibrillation, however, a proportion of patients may experience the major risk of bleeding, which can cause significant morbidity or mortality. Hence, it is possible that the majority of individuals have a reduced risk on stroke due to treatment while the treatment still causes death to a subgroup of them. The individual surrogate paradox manifests in this case. % since although the treatment is randomized, the surrogate and primary outcome may still be confounded by some unmeasured confounders. 

%is important since it sheds light on how to evaluate treatment for different individuals.  While traditional medical care selects one treatment that fits all, modern medical procedure tailors the medical interventions to each individual to accommodate the individuals heterogeneity (\citet{verma2012personalized,zhao2012estimating,zhao2013recent} and references therein). For example, wearable activity trackers such as Fitbit and smart phones are utilized to craft personalized health strategies \citep{dempsey2015randomised,kumar2013mobile}. This is known as the personalized treatment or precision medicine. However, not only the treatment needs to be personalized, how to evaluate treatment also needs to be customized. To the best of our knowledge, this has been rarely considered in the literature. Understanding how the surrogate endpoint serves as a substitute for each individual is important for the decision on who to treat, when to treat and how to treat.

The exclusion of individual surrogate paradox is challenging. First of all, inference on the proportion of individuals benefitted or get harmed by the treatment is in general harder than that of the average treatment effect since the former involves the joint distribution of potential outcomes \citep{yin2017assessing}. In randomization studies, the average treatment effect is easily obtained by comparing the average responses between treatment and control groups but the proportion of individuals benefitted or harmed by the treatment is still not identifiable \citep{gadbury2004individual}. Technically, if all the effect modifiers and confounders are collected and the mechanism of how treatment affects the surrogate and primary outcome is correctly modeled, then all the effects are homogeneous within the subgroups and the individual surrogate paradox could be avoided by making inference within subgroups. However, in a case where surrogate is typically used, the underlying mechanism of the disease is unknown. Thus, it is hard to pin point or even observe all the critical factors. Furthermore, even when both the ACE and DCE surrogate paradoxes are absent, the individual surrogate paradox can still manifest in both randomized trials and observational studies. 

In this paper, we propose the individual surrogate paradox. As mentioned above, a good surrogate should avoid the individual surrogate paradox. Hence, the investigation of the individual surrogate paradox provides an individual perspective to evaluate the surrogate variables. We investigate whether the individual surrogate paradox could manifest under the existing criteria. We found that among all the existing methods, only the strong binary surrogate can avoid the individual surrogate paradox without additional assumptions while other methods cannot exclude the paradox even for binary surrogate. Based on sharp bounds of the harm rate, we provide new criteria to exclude the paradox and incorporate a generalized causal necessity assumption to further boost power.

% \citep{rothwell2005subgroup}
%
%If we already know that the treatment does no harm to the surrogate and the surrogate does no harm to the true end point, it would contain far more information than the ACE of them are positive. Not only do we wonder the ACE of treatment on the outcome, but also want to know whether it is harmful to some specific people. For example, some certain drugs can reduce the level of blood sugar for any patient with hyperglycemia. At the same time, reducing blood sugar levels can lower the chance of cardiovascular diseases, which means if a cardiovascular disease does not occur in a higher level of glucose, it cannot occur in a lower level of glucose. We regard the drug as treatment, level of blood sugar as surrogate, and the cardiovascular disease as outcome. Therefore, the treatment harm rate (HR) of treatment on the surrogate, and the HR of surrogate on the outcome are all 0. At this time, we wonder not only the ACE of treatment on the outcome but also the treatment harm rate on the outcome. Moreover, pharmaceutical companies are keen to know the outcome for each person since it would cause large economic lost if the treatment has a negative causal effect for someone.
%
We organize the paper as follows. In Section \ref{sec: prelim}, we define individual surrogate paradox. In Section \ref{sec: existing}, we explore existing criteria on their ability to exclude the individual surrogate paradox. In Section \ref{sec: new_criteria}, we propose new criteria to exclude or confirm the presence of individual surrogate paradox, and in Section \ref{sec: data}, we demonstrate our criteria through data analysis. In Section \ref{sec: discuss}, we present a brief discussion.

\section{Preliminaries}\label{sec: prelim}

Let $T$ denote the randomized treatment variable, $Y$ the outcome and $S$ the surrogate. Although the treatment is randomized, the relationship between the surrogate and the primary outcome may in general still be confounded. Let $W$ denote the unmeasured confounder between the surrogate $S$ and the outcome $Y$. We restrict $T$ to be binary throughout the paper. Let $T=1$ denote the active treatment and $T=0$ denote the placebo. Let $S_{t}$ denote the potential outcome of surrogate if the treatment was set to $T=t$, and $Y_{ts}$ denote the potential outcome of the primary endpoint if the treatment and the surrogate were set to $T=t$ and $S=s$ by an external intervention on $T$ and $S$. We may also use the notation that $Y_{T=t}=Y_{tS_{t}}$ as the potential primary outcome when the intervention is only to set $T=t$. Without loss of generality, we assume for both $S$ and $Y$, that the larger values correspond to the better results.

Let $\mathrm{HR}(T\rightarrow S)=P(S_{0}-S_{1}>0)$ denote the harm rate of treatment $T$ on surrogate endpoint $S$, i.e., the proportion of individuals who have a worse surrogate endpoint under treatment versus control. Similarly, let $\mathrm{HR}(T\rightarrow Y)=P(Y_{0S_{0}}-Y_{1S_{1}}>0)$ denote the harm rate of treatment $T$ on primary outcome $Y$. Let $\mathrm{HR}_{s_{0},s_{1}}(S\rightarrow Y|T=t)=P(Y_{ts_{0}}-Y_{ts_{1}}>0)$ denote the harm rate of surrogate $S$ of two specific levels $s_0$, $s_1$, on primary endpoint $Y$ given treatment $T=t$. When surrogate $S$ is binary, we simplify the notation $\mathrm{HR}_{s_{0},s_{1}}(S\rightarrow Y|T=t)$ as $\mathrm{HR}(S\rightarrow Y|T=t)$. For a good surrogate, if the majority of individuals get a beneficial treatment effect on surrogate and a beneficial surrogate effect on the primary endpoint, then it is expected that the majority of individuals would get a beneficial treatment effect on the primary endpoint. Otherwise, paradoxical result happens on the individual level. Formally, we define the individual surrogate paradox as below for any given constants $0\leq c_1\leq 1$ and $0\leq c_2\leq 1$.

\vspace{-5mm}
\begin{defn}\label{def: disparity paradox}
Under the conditions that (i) the proportion of individuals who have a worse surrogate endpoint under the treatment $T$ is no greater than $c_1$, i.e., $\mathrm{HR}(T\rightarrow S)=P(S_{0}-S_{1}>0)\leq c_1$, and (ii) the proportion of individuals that surrogate $S$ does harm to primary outcome $Y$ is no greater than $c_2$ , i.e., $\mathrm{HR}_{s_0,s_1}(S\rightarrow Y|T=t)=P(Y_{ts_{0}}-Y_{ts_{1}}>0)\leq c_2$ for any $s_{1}>s_{0}$ and $t=0,1$, then, the individual surrogate paradox is present if (iii) the proportion of individuals that treatment $T$ does harm to primary outcome $Y$ is greater than $c_1+c_2$, i.e., $\mathrm{HR}(T\rightarrow Y)=P(Y_{0S_{0}}-Y_{1S_{1}}>0)>c_1+c_2$.
\end{defn}

\vspace{-3mm}
In Definition \ref{def: disparity paradox} (iii), the threshold of $\mathrm{HR}(T\rightarrow Y)$  is chosen as $c_1+c_2$, however, this threshold can be generalized to any number $0\leq \tilde c\leq 1$, and our proposed methods can be generalized easily. For the discussion in this paper, we only focus on the individual surrogate paradox with threshold $c_1+c_2$ for $\mathrm{HR}(T\rightarrow Y)$. That is, the individual surrogate paradox manifests if there are more individuals get harmful treatment effect on primary outcome than the total number of individuals who have a harmful treatment effect on surrogate and of those who have a harmful surrogate effect on primary outcome. 

The surrogate paradox given in Definition \ref{def: disparity paradox} directly compares the two potential outcomes (e.g., $Y_0$ and $Y_1$) of the same individual and focuses on the scenario that one is better than the other. In contrast, \citet{chen2007criteria} defined the surrogate paradox based on the average causal effect and \citet{ju2010criteria} defined that based on difference between distributions, both of which are population measures. At the end of this section, we will show that even when both the ACE and the DCE surrogate paradox are avoided, the individual surrogate paradox can still manifest.

%\citet{chen2007criteria} defined the average causal effect of treatment $T$ on primary outcome $Y$ as $ACE(T\xrightarrow{}Y)=E(Y_1-Y_0)$ and the ACE surrogate paradox is 
%
%\begin{itemize}
%  \item \citep{chen2007criteria} A treatment has a positive ACE on a surrogate which in turn has a positive ACE on a primary outcome, but the treatment has a negative ACE on the primary outcome, i.e., $\mathrm{ACE}(T\rightarrow S)>0$ and $\mathrm{ACE}(S\rightarrow Y)>0$, but $\mathrm{ACE}(T\rightarrow Y)<0$.
%\end{itemize}
%
%\noindent  \citet{ju2010criteria} 
Since the harm rate $\mathrm{HR}(T\rightarrow Y)$ is between 0 and 1, we further restrict the range of $c_2$ as $0\leq c_2\leq1-c_1$. A special case of the individual surrogate paradox is when $c_1=c_2=0$, that is, for everyone, neither the treatment does any harm to the surrogate endpoint nor the surrogate does any harm to the primary endpoint, but the treatment does harm to the primary outcome for some people. This special case has a nice feature that if a surrogate does not suffer from the individual surrogate paradox for a population, then it does not suffer from such paradox for any the subpopulations. Hence, whether the individual surrogate paradox manifests can be used as a new criterion for evaluating surrogate endpoint and such criterion is consistent for all the subpopulations. Equivalently, under this criterion, if a surrogate is not a good surrogate for a subgroup, then it is not a good surrogate for any bigger population. In comparison, a good surrogate in terms of avoiding either the ACE or DCE surrogate paradox  \citep{chen2007criteria,ju2010criteria} on one population does not have any indication of whether this is a good surrogate on any subpopulations. Furthermore, if the individual surrogate paradox with $c_1=c_2=0$ is excluded, then the surrogate paradox based on ACE is guaranteed to be absent as well. 

Now, we give an example that even when both the ACE and the DCE surrogate paradoxes are absent, the individual surrogate paradox can still manifest.  For simplicity, we construct an example where surrogate $S$ and outcome $Y$ are both binary. In the binary case, DCE and ACE are equivalent. As a non-negative individual causal effect for everyone implies a non-negative average causal effect, we only need to construct an example such that  (1) treatment has a non-negative individual causal effect on surrogate, (2) surrogate has a non-negative individual causal effect on primary outcome, and (3) on average, the treatment has a non-negative ACE on primary outcome, but (4) for some individuals, treatment has a negative individual causal effect on the primary outcome. To construct such example, we need to specify the joint probabilities of potential outcomes. There are 4 potential outcomes  for the primary endpoint $Y_{ts}$ for $t=0,1$ and $s=0,1$ and 16 different possible values of the vector $(Y_{00},Y_{01},Y_{10},Y_{11})$. Additionally, there are 2 potential outcomes for surrogate $S_{t}$ for $ t=0,1,$ and 4 different possible values for $(S_{0},S_{1})$. Thus, we have $16\times4=64$ different possible values for the vector $(Y_{00},Y_{01},Y_{10},Y_{11},S_{0},S_{1})$. For ease of illustration, let $q_{i,j}\geq0$ denote the proportion of getting each possible value of $(Y_{00},Y_{01},Y_{10},Y_{11},S_{0},S_{1})$ in the whole population for $i=0,\ldots,15$ and $j=0,\ldots,3$ (shown in Table \ref{tb: non_strong_q} in the Supplementary Materials). Under conditions (i) and (ii) in Definition \ref{def: disparity paradox} with $c_1=c_2=0$, we can reduce the 64 categories to 27 categories since the others are all 0 as shown in Table \ref{tb: nonstrong_q_noindi_para}. For example, since $\mathrm{HR}(T\rightarrow S)=P(S_1=0, S_0=1)=\sum_{i=0}^{15}q_{i,2}=0$,  we have $q_{0,2}=q_{1,2}=\ldots=q_{15,2}=0$. %Similarly, we have  %$\mathrm{HR}(S\rightarrow Y|T=0)=P(Y_{01}=0, Y_{00}=1)=\sum_{j=0}^{3}\sum_{i=8}^{11}q_{i,j}=0$ and $\mathrm{HR}(S\rightarrow Y|T=1)=P(Y_{11}=0, Y_{10}=1)=\sum_{j=0}^{3}\sum_{i=2,6,10,14}q_{i,j}=0$, and therefore, 
 %$q_{8,0}=\ldots=q_{8,3}=q_{9,0}=\ldots=q_{11,3}=0$ and $q_{2,0}=\ldots=q_{2,3}=q_{6,0}=\ldots=q_{10,3}=0$. 
 The following example demonstrates that the absence of surrogate paradoxes based on ACE and DCE cannot exclude the existence of individual surrogate paradox.

 \begin{example}
 	  When the surrogate $S$ and the outcome $Y$ are both binary, the probability of $q_{i,j}$ is given in Table \ref{tb: eg_prentice}, then, the surrogate paradoxes based on ACE and DCE are both absent while individual surrogate paradox manifests.
 	 \label{example1}
 \end{example}
 
 \vspace{-4mm}
We have $\mathrm{HR}(T\rightarrow S)=0$, and $\mathrm{HR}(S\rightarrow Y|T=t)=0$ for $t=0,1$, but $\mathrm{HR}(T\rightarrow Y)=P(Y_{T=0}=1,Y_{T=1}=0)=q_{4,3}+q_{12,0}+q_{12,1}+q_{12,3}+q_{13,0}=0.235>0$. Additionally, we have $\mathrm{ACE}(T\rightarrow Y)=0.03>0$. Hence, both the surrogate paradoxes based on ACE and DCE are absent but the individual surrogate paradox manifests. As mentioned, both surrogate paradoxes based on ACE and DCE are based on population measures, while the individual surrogate paradox focuses on the individual performance. Thus, it is not surprising that the absence of the formers cannot exclude the presence of the latter.

%We first show that both the surrogate paradoxes based on ACE and DCE are absent. Specifically, we obtain that $\mathrm{ACE}(T\rightarrow S)=0.4>0$, $\mathrm{ACE}(S\rightarrow Y|T=0)=0.225>0$, $\mathrm{ACE}(S\rightarrow Y|T=1)=0.27>0$, $\mathrm{ACE}(T\rightarrow Y)=0.03>0$. Hence, the surrogate paradox based on ACE does not occur. Besides, when $T$, $S$, $Y$ are binary, DCE is equivalent to ACE. Thus, surrogate paradox based on DCE can also be avoided. On the other hand, we have $\mathrm{HR}(T\rightarrow S)=0$, and $\mathrm{HR}(S\rightarrow Y|T=t)=0$ for $t=0,1$, but $\mathrm{HR}(T\rightarrow Y)=P(Y_{T=0}=1,Y_{T=1}=0)=q_{4,3}+q_{12,0}+q_{12,1}+q_{12,3}+q_{13,0}=0.235>0.$ This means $T$ does harm to $Y$ for 23.5\% of the individuals even when no one gets harmed by the treatment effect on surrogate and the surrogate effect on the primary outcome. Hence, the individual surrogate paradox manifests.

%Various criteria have been proposed to evaluate the surrogate endpoint. In the next section, we investigate whether individual surrogate paradox would manifest under the existing surrogate criteria. 

%The surrogate paradox raised by \citet{chen2007criteria} suggested the ACE of $T$ on $S$ and the ACE of $S$ on $Y$ cannot accurately predict the ACE of $T$ on $S$, even the sign. The individual surrogate paradox indicates that $T$ does little harm to
%$S$, and $S$ does little harm to $Y$ cannot guarantee $T$ does little harm to $Y$. Here do little harm simply means $\mathrm{HR}(S\rightarrow Y|T=t)$ is no greater than a very small number. 

\section{Existing Criteria and Individual Surrogate Paradox}\label{sec: existing}

\subsection{Prentice criterion}\label{subsec: prentice}

Prentice criterion requires that when given surrogate $S$, treatment $T$ is independent of outcome $Y$, that is, $T\bot Y|S$ \citep{prentice1989surrogate}. The intuition of this criterion is to have the surrogate blocks all the association between treatment and primary endpoint. The Prentice criterion is defined based on population distribution, but the individual surrogate paradox is defined from the perspective of individual. So intuitively, Prentice criterion cannot guarantee the exclusion of this paradox. 

Example \ref{example1}  also serves as an counterexample with binary surrogate and outcome that Prentice criterion cannot exclude the individual surrogate paradox when $c_1=c_2=0$. (See Supplementary materials for another counterexample with continuous surrogate $S$ and outcome $Y$). For Example \ref{example1}, we have already demonstrate that the individual surrogate paradox manifests. To verify the Prentice criterion, we have
$$\left\{
\begin{array}{ll}
P(Y=1|T=0,S=0)
=\dfrac{q_{12,0}+q_{13,0}+q_{15,0}+q_{12,1}+q_{13,1}+q_{15,1}}{\sum_{i}(q_{i,0}+q_{i,1})}=0.5   ,  \\
P(Y=1|T=1,S=0)
=\dfrac{q_{3,0}+q_{7,0}+q_{15,0}}{\sum_{i}q_{i,0}}=0.5   ,  \\
P(Y=1|T=0,S=1)
=\dfrac{q_{4,3}+q_{5,3}+q_{7,3}+q_{12,3}+q_{13,3}+q_{15,3}}{\sum_{i}q_{i,3}}=0.75   ,  \\
P(Y=1|T=1,S=1)
=\dfrac{\sum_{j=1,3}q_{1,j}+q_{3,j}+q_{5,j}+q_{7,j}+q_{13,j}+q_{15,j}}{\sum_{i}(q_{i,1}+q_{i,3})}=0.75.
\end{array}
\right.$$

\noindent Thus, we have $P(Y=1|T=0,S=s)=P(Y=1|T=1,S=s)$ for $s=0,1$ and the Prentice criterion holds, i.e.,  $Y\bot T|S$.

\subsection{Principle surrogate criterion} 

Principle surrogate requires $Y_{T=0}$ and $Y_{T=1}$ to be identically distributed under the principle strata $ S_{0}=S_{1} $, i.e., $P(Y_{T=0}|S_{0}=S_{1}=s)=P(Y_{T=1}|S_{0}=S_{1}=s),$ for all $s$ \citep{frangakis2002principal}. This criterion is based on the distribution of each potential outcomes in this strata rather than comparing the two potential outcomes of the same individual directly. So intuitively, principle surrogate cannot avoid the paradox. We present one counterexample below when $c_1=c_2=0$ for binary surrogate and outcome. Another counterexample of continuous variables are given in the Supplementary Materials.

\begin{example}
	When surrogate $S$ and outcome $Y$ are both binary, the probabilities of the joint potential outcomes  $q_{i,j}$ are given in Table \ref{counter_principal_bin} in the Supplementary Materials. Then, the principle surrogate criterion holds while individual surrogate paradox manifests.
	\label{example3}
\end{example}

Specifically, we have

$$\left\{
\begin{array}{ll}
P(Y_{T=0}=1|S_{0}=S_{1}=0)=\dfrac{q_{12,0}+q_{13,0}+q_{15,0}}{\sum_{i}q_{i,0}}=1/3  ,  \\
P(Y_{T=1}=1|S_{0}=S_{1}=0)=\dfrac{q_{3,0}+q_{7,0}+q_{15,0}}{\sum_{i}q_{i,0}}=1/3  ,  \\
P(Y_{T=0}=1|S_{0}=S_{1}=1)=\dfrac{q_{4,3}+q_{5,3}+q_{7,3}+q_{12,3}+q_{13,3}+q_{15,3}}{\sum_{i}q_{i,3}}=3/10  ,  \\
P(Y_{T=1}=1|S_{0}=S_{1}=1)=\dfrac{q_{1,3}+q_{3,3}+q_{5,3}+q_{7,3}+q_{13,3}+q_{15,3}}{\sum_{i}q_{i,3}}=3/10  .  
\end{array}
\right.$$

\noindent Therefore, $P(Y_{T=0}=1|S_{0}=S_{1}=s)=P(Y_{T=1}=1|S_{0}=S_{1}=s)$ for $s=0,1$, indicating $S$ is a principle surrogate. On the other hand, $\mathrm{HR}(T\rightarrow S)=\mathrm{HR}(S\rightarrow Y|T=t)=0$ for $t=0$, 1, and $\mathrm{HR}(T\rightarrow Y)=P(Y_{T=0}=1,Y_{T=1}=0)=q_{4,3}+q_{12,0}+q_{12,1}+q_{12,3}+q_{13,0}=0.235>0.$
Hence, the individual surrogate paradox manifests.

\subsection{Strong Surrogate Criterion}
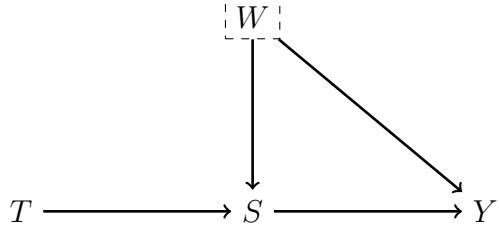
\begin{figure}
	\centering
	\begin{tikzpicture}
	% nodes %
	\node[text centered] (t) {$T$};
	\node[right = 2.5 of t, text centered] (s) {$S$};
	\node[right=2.5 of s, text centered] (y) {$Y$};
	\node[draw, rectangle, dashed, above = 2 of s, text centered] (u) {$W$};
	
	% edges %
	\draw[->, line width= 1] (t) --  (s);
	\draw [->, line width= 1] (s) -- (y);
	\draw[->,line width= 1] (u) --(s);
	\draw[->,line width= 1] (u) -- (y);
	\end{tikzpicture}
	\medskip
	\caption{Causal diagram of the strong surrogate $S$ for the effect of the treatment $T$ on outcome $Y$\label{DAGstr}}
\end{figure}

%To easy the illustration, we first focus on the strong surrogate case. 
\citet{lauritzen2004discussion} defined a surrogate $S$ to be a strong surrogate when there is no direct path between treatment $T$ and primary outcome $Y$ without going through surrogate $S$. (Figure \ref{DAGstr}). Under this condition, potential outcome $Y_{ts}$ can be simply written as $Y_{s}$. Hence, we have $\mathrm{HR}_{s_0,s_1}(S\rightarrow Y|T=t)=P(Y_{s_1}-Y_{s_0}>0)$, which we simplify as $\mathrm{HR}_{s_0,s_1}(S\rightarrow Y)$ and further simplify as $\mathrm{HR}(S\rightarrow Y)$ for binary surrogate $S$. \citet{chen2007criteria} showed that the strong surrogate cannot avoid the surrogate paradox based on ACE. We demonstrate that a strong binary surrogate can avoid the individual surrogate paradox without any additional assumptions. To see this, note we have that%Therefore, we use $Y_{s}$ denote the potential outcome given $S=s$.
\begin{eqnarray*}
%	&&\mathrm{HR}(T\rightarrow Y)\\%= P(Y_{T=0,S_{0}}-Y_{T=1,S_{1}}>0) \\
	%&=&P(Y_{T=0}-Y_{T=1}>0) \\
	%&=& P(Y_{T=0,S_{0}}-Y_{T=1,S_{1}}>0) \\
	\mathrm{HR}(T\rightarrow Y)&=& P(Y_{S_{0}}-Y_{S_{1}}>0) \\
%	&=& P(Y_{S_{0}}-Y_{S_{1}}>0,S_{1}>S_{0})+P(Y_{S_{0}}-Y_{S_{1}}>0|S_{1}<S_{0})P(S_{1}<S_{0}) \\
	&=& P(Y_{S_{0}=0}-Y_{S_{1}=1}>0,S_{1}=1,S_{0}=0)+P(Y_{S_{0}}-Y_{S_{1}}>0|S_{1}<S_{0})P(S_{1}<S_{0}) \\
	&=&P(Y_{S=0}-Y_{S=1}>0)P(S_{1}=1,S_{0}=0|Y_{S=0}-Y_{S=1}>0)\\
	&&+P(Y_{S_{0}}-Y_{S_{1}}>0|S_{1}<S_{0})P(S_{1}<S_{0}) \\
	&\leq&P(Y_{S=0}-Y_{S=1}>0)+P(S_{1}<S_{0}) \leq c_1+c_2,
\end{eqnarray*}
where the first equation is due to the strong surrogate assumption, and the second equation is obtained by the total probability theorem. That is, strong binary surrogate will never encounter the individual surrogate paradox.  This makes a nice feature of the strong surrogate.

Unfortunately, for non-binary surrogate, even the strong surrogate criterion cannot exclude the individual surrogate paradox. In the Supplementary Material, we present a counterexample that when $S$ has three levels of possible values, under strong surrogate condition, individual surrogate paradox may still manifest.

\subsection{Wu and VanderWeele Criteria}
Let $f(s,t)$ denote the expectation $E(Y|S=s,T =t)$ of $Y$ conditional on $S=s$ and $T =t$. \citet{wu2011sufficient} proposed the following criterion to exclude the surrogate paradoxes based on ACE and DCE.

%\begin{nono-crit}[\bf Wu et al., 2011\nocite{wu2011sufficient}]
\vspace{2mm}
\textsc{Criterion} \textbf{(Wu et al., 2011\nocite{wu2011sufficient}):}	A surrogate S should satisfies the following conditions:
	
	1. $f (s,1)$ or $f (s,0)$ monotonically increases as $s$ increases (i.e. $f(s',1)\geq f (s'',1)$ for all $s'>s''$ or $f (s',0)\geq f (s'',0)$ for all $s'>s''$), and
	
	2. $f(s,1)\geq  f(s,0)$ for all $s$.
\vspace{2mm}
	
%\noindent	Then $T$ has a non-negative ACE on $Y$ if $T$ has a non-negative DCE on $S$.
%\end{nono-crit}

This criterion is also based on the observed data distribution, so intuitively, it cannot exclude the individual surrogate paradox. We show in the Supplementary material that Example \ref{example1} serves as a counterexample that Wu criterion cannot exclude the individual surrogate paradox when $S$ and $Y$ are both binary. 

%\begin{example}
%	The probability of $q_{i,j}$ is given in Table \ref{tb: eg_prentice}. Since this example is the same as Example \ref{example1}, we have already indicate individual surrogate paradox manifests. What we will show below is the criteria proposed by \citep{wu2011sufficient} is satisfied.
%\end{example}
%From Table \ref{tb: eg_prentice}, we have $f(1,1)=P(Y=1|S=1,T=1)=0.75$, $f(1,0)=P(Y=1|S=1,T=0)=0.75$, $f(0,1)=P(Y=1|S=0,T=1)=0.5$, and $f(0,0)=P(Y=1|S=0,T=0)=0.5$. Hence, we have $f(1,1)\geq f(0,1)$, $f(1,0)\geq f(0,0)$, $f(0,1)=f(0,0)$, $f(1,1)=f(1,0)$ satisfying the condition of \citet{wu2011sufficient}. However, we have $\mathrm{HR(T\rightarrow Y)}=0.235>0$. Thus, individual surrogate paradox manifests.
%

\citet{vanderweele2013surrogate} provided the following sufficient conditions to avoid the surrogate paradox with a non-strong surrogate.

%\begin{nono-crit}[\bf VanderWeele, 2013\nocite{vanderweele2013surrogate}]\label{crit: vander}
\vspace{2mm}
\textsc{Criterion} \textbf{(VanderWeele, 2013\nocite{vanderweele2013surrogate}):}
A surrogate should satisfies the following conditions: $ (a) $ $ E(Y|t,s,w) $ is non-decreasing in $ t $ and $ s $ for all $ w $ and $ (b) $ $ P(S > s|t, w) $ is non-decreasing in $ t $ for all $ s, w $. %, then $ E(Y_{t})  = E(Y|t) $ is non-decreasing in $ t $.	
%\end{nono-crit} 
\vspace{2mm}

VanderWeele criterion is also based on population measures, thus, it cannot exclude the individual surrogate paradox (see supplementary material).
 Since the surrogate criterion of \citet{vanderweele2013surrogate} is more general than those in \citet{chen2007criteria} and \citet{ju2010criteria}, counterexample for VanderWeele criterion also serves as an counterexample to show that criteria in \citet{chen2007criteria} and \citet{ju2010criteria} cannot exclude the individual surrogate paradox.

%When both $S$ and $Y$ are binary, we have $\mathrm{HR}(T\rightarrow Y)=P(Y_{T=0}=1,Y_{T=1}=0)=q_{4,3}+q_{12,0}+q_{12,1}+q_{12,3}+q_{13,0}.$
%Assume $W$ is binary with $ P(W=1)=P(W=0)=0.5$ and the joint distribution of $(S_0,S_1,Y_{00},Y_{01}, Y_{10},Y_{11})|W=w$ the same as the probabilities in given in Table \ref{tb: eg_prentice} for $w=0,1$. For example, assume $P(Y_{00}=1, S_0=0|W=w)=P(Y_{00}=1, S_0=0)= 0.3$. Then, we have
%$$\left\{
%\begin{array}{ll}
%P(Y=1|T=0,S=0,W=w)=\dfrac{P(Y_{00}=1,S_{0}=0|W=w)}{P(S_{0}=0|W=w)}
%=0.5   ,  \\
%P(Y=1|T=1,S=0,W=w)=\dfrac{P(Y_{10}=1,S_{1}=0|W=w)}{P(S_{1}=0|W=w)}
%=0.5   ,  \\
%P(Y=1|T=0,S=1,W=w)=\dfrac{P(Y_{01}=1,S_{0}=1|W=w)}{P(S_{0}=1|W=w)}
%=0.75   ,  \\
%P(Y=1|T=1,S=1,W=w)=\dfrac{P(Y_{11}=1,S_{1}=1|W=w)}{P(S_{1}=1|W=w)}
%=0.75   .
%\end{array}
%\right.$$
%Therefore, condition (a) is satisfied. Additionally, we have $P(S=1|T=0,w)=0.4\leq 0.8=P(S=1|T=1,w)$ satisfying condition (b).
%Since $\mathrm{HR}(T\rightarrow Y)=P(Y_{T=0}=1,Y_{T=1}=0)=q_{4,3}+q_{12,0}+q_{12,1}+q_{12,3}+q_{13,0}=0.235>0$, $T$ does harm to $Y$ for 23.5\% of the individuals. %Individual surrogate paradox happens in this example.
%

\section{New Criteria for the Individual Surrogate Paradox}\label{sec: new_criteria}

\begin{figure}
	\centering
	\begin{tikzpicture}
	% nodes %
	\node[text centered] (t) {$T$};
	\node[right = 2.5 of t, text centered] (s) {$S$};
	\node[right=2.5 of s, text centered] (y) {$Y$};
	\node[draw, rectangle, dashed, above = 2 of s, text centered] (w) {$W$};
	
	% edges %
	\draw[->, line width= 1] (t) --  (s);
	\draw [->, line width= 1] (s) -- (y);
	\draw[->,line width= 1] (w) --(s);
	\draw[->,line width= 1] (w) -- (y);
	\draw[->,line width=1] (t) to  [out=270,in=270, looseness=0.5] (y);
	\end{tikzpicture}
	\medskip
	\caption{Causal diagram of the strong surrogate $S$ for the effect of the treatment $T$ on outcome $Y$\label{DAGnon}}
\end{figure}
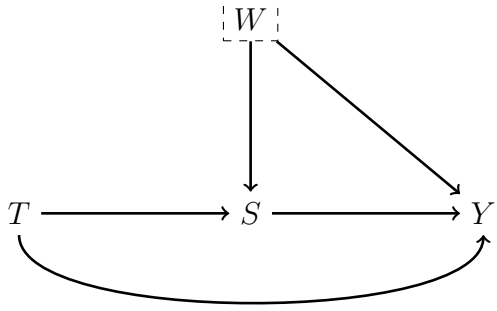

In this section, we propose several criteria to exclude as well as to confirm the presence of the individual surrogate paradox. As discussed, if a binary surrogate is a strong surrogate, then the individual surrogate paradox will not manifest.  Unfortunately, the strong surrogate assumption may be less plausible in some scenarios. For example, when evaluating the effectiveness of chemotherapy, 6-months progression status can be used as surrogate endpoint for overall survival, however, the chemotherapy can also affect the survival via side effects such as damaging cells in the heart, kidneys, bladder, lungs, and nervous system. Hence, the 6-months progression status is  a non-strong surrogate. That is, treatment $T$ can directly have an effect on outcome $Y$ instead of going through surrogate $S$ (see Figure \ref{DAGnon}). Additionally, it is also of interest to exclude as well as to confirm the presence of the individual surrogate paradox for non-binary surrogate.

We first give a necessary condition for the presence of individual surrogate paradox. In Example \ref{example1}, we have $P(Y_{T=1,S_{1}}-Y_{T=0,S_{1}}<0)=0.25>0$. Actually, it can be shown that $P(Y_{T=1,S_{1}}-Y_{T=0,S_{1}}<0)>0$ is a necessary condition for the individual surrogate paradox to happen with $c_1=c_2=0$. Note we have
\begin{eqnarray*}
	P(Y_{T=0}-Y_{T=1}>0) &=& P(Y_{T=0,S_{0}}-Y_{T=1,S_{1}}>0) \\
	&=& P(Y_{T=0,S_{0}}-Y_{T=1,S_{1}}>0|S_{1}\geq S_{0}) \\
	&=& P(Y_{T=0,S_{0}}-Y_{T=0,S_{1}}+Y_{T=0,S_{1}}-Y_{T=1,S_{1}}>0|S_{1}\geq S_{0}) \\
	&=& P(Y_{T=0,S_{1}}-Y_{T=1,S_{1}}>Y_{T=0,S_{1}}-Y_{T=0,S_{0}}|S_{1}\geq S_{0})  \\
	&\leq& P(Y_{T=0,S_{1}}-Y_{T=1,S_{1}}>0|S_{1}\geq S_{0})  \\
	&=& P(Y_{T=1,S_{1}}-Y_{T=0,S_{1}}<0|S_{1}\geq S_{0})  \\
	&=& P(Y_{T=1,S_{1}}-Y_{T=0,S_{1}}<0). 
\end{eqnarray*}

The second and the last equation come from the fact that  $T$ does no harm to $S$, i.e., $P(S_{1}\geq S_{0})=1$. The inequality holds because $S$ does no harm to $Y$, i.e., $\mathrm{HR}(S\rightarrow Y|T=t)=P(Y_{T=0,s_{1}}-Y_{T=0,s_{0}}<0)=0$ for any $s_1>s_0$ and hence $P(Y_{T=0,S_{1}}-Y_{T=0,S_{0}}<0|S_1>S_0)=0$. The above derivation indicates that only when there exist individuals that the direct causal effect from $T$ to $Y$ is negative, can we obtain $P(Y_{T=0}-Y_{T=1}>0) >0$, which leads to individual surrogate paradox. Hence, the condition $P(Y_{T=1,S_{1}}-Y_{T=0,S_{1}}<0)=0$ can be used as a sufficient condition for the exclusion of individual surrogate paradox. However, it is hard to verify this using the observed data. Below, we explore more practical criteria for excluding the individual surrogate paradox. %Surrogate paradox occurs due to the direct negative causal effect from $T$ to $Y$.  

For ease of illustration, we first only focus on the case where both surrogate and primary outcome are binary.  To exclude or confirm the presence of individual surrogate paradox, we first derive the sharp bound of $\mathrm{HR}(T\rightarrow Y)$ under the condition that the proportion of individuals that have a negative treatment effect on surrogate is no greater than $c_1$, i.e., $\mathrm{HR}(T\rightarrow S)\leq c_1$ and the proportion of individuals that have a negative surrogate effect on the primary outcome is no greater than $c_2$ for both treatment arms, i.e., $\mathrm{HR}(S\rightarrow Y|T=t)=P(Y_{t0}=1, Y_{t1}=0)\leq c_2$ for $t=0$, $1$. We have the following Proposition for binary surrogate $S$ and outcome $Y$.

\begin{prop}\label{prop: non-strong}
Given $P(Y,S|T)$, suppose $\mathrm{HR}(T\rightarrow S)\leq c_1$ and $\mathrm{HR}(S\rightarrow Y|T=t)\leq c_2$ for $t=0$, $1$, then, the sharp bound of $\mathrm{HR}(T\rightarrow Y)$ is $L\leq \mathrm{HR}(T\rightarrow Y)\leq U,$
where $$L=\max\left(
\begin{array}{c}
L_{1}  \\
L_{2}  \\
L_{3}  \\
L_{4}  \\
\end{array}
\right)=\max\left(
\begin{array}{c}
0  \\
P(Y=1,S=1|T=0)-P(Y=1,S=1|T=1)-c_{1}  \\
P(Y=0|T=1)-P(Y=0|T=0)  \\
P(Y=0,S=0|T=1)-P(Y=0,S=0|T=0)-c_{1}  \\
\end{array}
\right),
$$

$$U_{}=\min\left(
\begin{array}{c}
U_{1}  \\
U_{2}  \\
U_{3}  \\
U_{4}  \\
\end{array}
\right)=\min\left(
\begin{array}{c}
P(Y=1|T=0)  \\
c_{1}+P(Y=1,S=0|T=0)+P(Y=0,S=1|T=1)  \\
P(Y=0|T=1)  \\
1+c_{1}-P(Y=0,S=1|T=0)-P(Y=1,S=0|T=1)  \\
\end{array}
\right).$$
\end{prop}

Bounds in Proposition \ref{prop: non-strong} are obtained using linear programming by considering the joint distributions of potential outcomes. Since both $L$ and $U$ only involve observed data distribution, they can be calculated using the available data. An observation yields that $L_{1}$, $L_{3}$, $U_{1}$, $U_{3}$ are simple bounds, in the sense that one can derive them without the conditions that $\mathrm{HR}(T\rightarrow S)\leq c_1$ and $\mathrm{HR}(S\rightarrow Y|T=t)\leq c_2$ for $t=0,1$. For example, we have  $\mathrm{HR}(T\rightarrow Y)=P(Y_{T=1}=0, Y_{T=0}=1)\geq P(Y_{T=1}=0)+P(Y_{T=0}=1)-1=P(Y=0|T=1)-P(Y=0|T=0)=L_{3}$.

Surprisingly, once given $P(Y,S|T)$ and $\mathrm{HR}(T\rightarrow S)\leq c_1$, the sharp bound of $\mathrm{HR}(T\rightarrow Y)$ does not dependent on $c_{2}$. That is, given $P(Y,S|T)$ and $\mathrm{HR}(T\rightarrow S)\leq c_1$, $\mathrm{HR}(T\rightarrow Y)$ does not change whether or not we collect $\mathrm{HR}(S\rightarrow Y|T=t)$. To see this, note that $\mathrm{HR}(T\rightarrow Y)=P(Y_{0S_{0}}=1, Y_{1S_{1}}=0)$, and $\mathrm{HR}(S\rightarrow Y|T=t)=P(Y_{t0}=1, Y_{t1}=0)$. That is, $\mathrm{HR}(T\rightarrow Y)$ involves the joint distribution of potential outcomes with different treatment values, but $\mathrm{HR}(S\rightarrow Y|T=t)$ involve the joint distribution of potential outcomes of the same treatment value. Hence, intuitively, the bounds of $\mathrm{HR}(T\rightarrow Y)$ is free of $c_2$. We present a rigorous proof in the Supplementary Materials. Since the sharp bound does not depend on $c_{2}$, it is reasonable to modify the definition of individual surrogate paradox as follows.
\begin{defn}\label{def: c1_only}
	
	\textit{The individual surrogate paradox is present if the treatment harm rate of $T$ on $S$ is no greater than $c_{1}$, i.e., $\mathrm{HR}(T\rightarrow S)=P(S_{0}=1, S_{1}=0)\leq c_{1}$, however, the treatment harm rate of $T$ on $Y$ is greater than $c_{1}$, i.e., $\mathrm{HR}(T\rightarrow Y)=P(Y_{0S_{0}}=1, Y_{1S_{1}}=0)>c_{1}$.}
\end{defn}

According to Definition \ref{def: c1_only}, the individual surrogate paradox manifests if there are more individuals that have a harmful treatment effect on primary outcome as compared with those have a harmful treatment effect on surrogate. Based on Proposition \ref{prop: non-strong}, we propose a criterion to exclude or conclude the presence of individual surrogate paradox when $S$ and $Y$ are both binary.
\begin{crit}\label{crit: non-strong} 

	(1) The individual surrogate paradox can be excluded if $U_{}=\min(U_{1}, U_{2},U_{3},U_{4})\leq c_{1}$;
	(2) the individual surrogate paradox is guaranteed to manifest if $L=\max(L_{1},L_{2},L_{3},L_{4})> c_1.$
\end{crit}

%CITET YIN SURROGATE PAPER. (1) is sufficient and almost necessary, (2) is sufficient. 
Criterion \ref{crit: non-strong} provides a feasible way to assess the presence of individual surrogate paradox as both $U$ and $L$ only involve observed data. As we show in the Supplementary Materials, the bounds proposed in Proposition \ref{prop: non-strong} are sharp. As a consequence, the conditions (1) and (2) in Criterion \ref{crit: non-strong} are both sufficient and ``almost necessary" \citep{yin2017optimal} to exclude or conclude the presence of individual surrogate paradox. That is, if (1) (or (2)) is satisfied, the individual surrogate paradox is guaranteed to be absent (or present); if (1) (or (2)) is not satisfied, then there exists another data generating mechanism, or joint density, that has (or does not have) the individual surrogate paradox and yields the same observed data. That is, Criterion \ref{crit: non-strong} has extracted the most information from the observed data with regards to the existence of individual surrogate paradox. A more thorough discussion of the sufficient and almost necessary property could be found in \citet{yin2017optimal}. 

As mentioned above, upper bounds $U_{1}$, $U_{3}$ are simple bounds and the upper bound $U$ is non-trivial when $U_{{2}}\leq \min(U_{{1}},U_{{3}})$ or $U_{{4}}\leq \min(U_{{1}},U_{{3}})$. This is equivalent to one of the two sets of the following inequalities holds:

$$\left\{
\begin{array}{ll} 
P(Y=1,S=0|T=0)<P(Y=0,S=0|T=1),\\ 
P(Y=0,S=1|T=1)<P(Y=1,S=1|T=0);
\end{array}
\right.$$

%\noindent or

$$\left\{
\begin{array}{ll} 
P(Y=1,S=1|T=1)<P(Y=0,S=1|T=0),\\ 
P(Y=0,S=0|T=0)<P(Y=1,S=0|T=1).
\end{array}
\right.$$

\begin{figure}
	\centering    
	\subfigure[Scenario 1]{\label{fig:U1}\includegraphics[width=50mm]{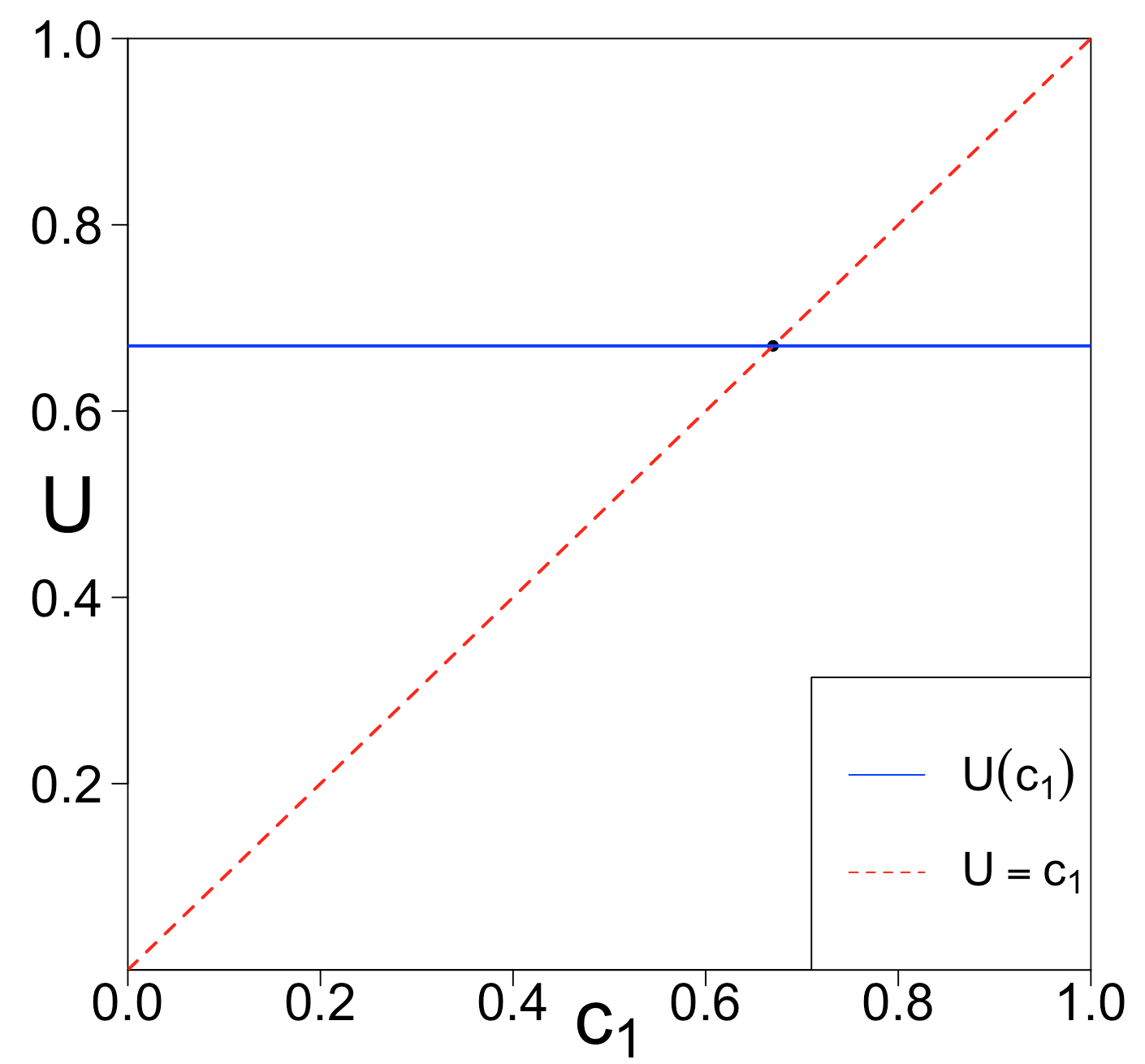}}
	\subfigure[Scenario 2]{\label{fig:U2}\includegraphics[width=50mm]{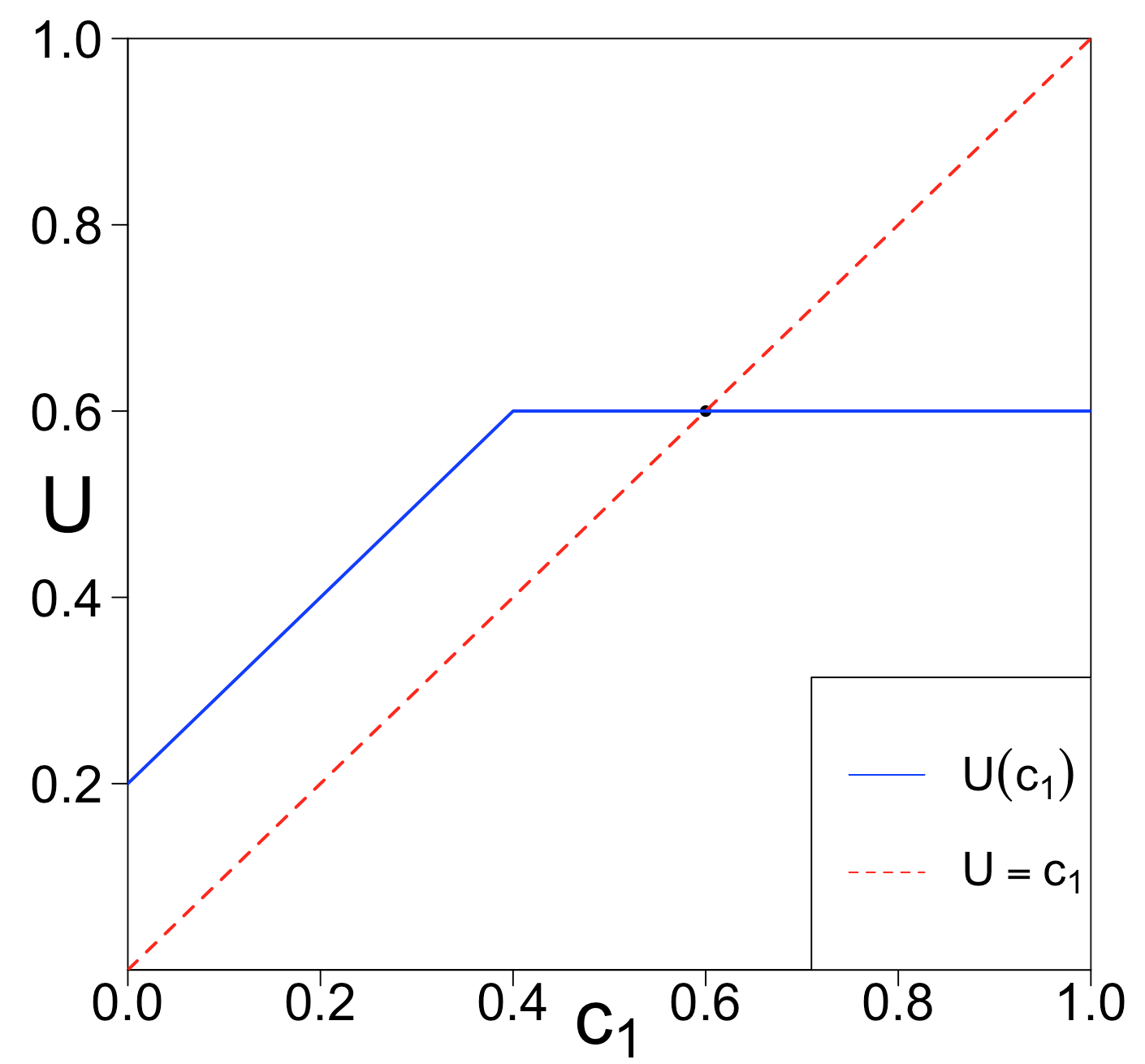}}
%	\caption{Two possible scenarios of piecewise linear functions for the upper bound $U$}
%	\label{fig:piecewise}
	
	\subfigure[Scenario 1]{\label{fig:L1}\includegraphics[width=50mm]{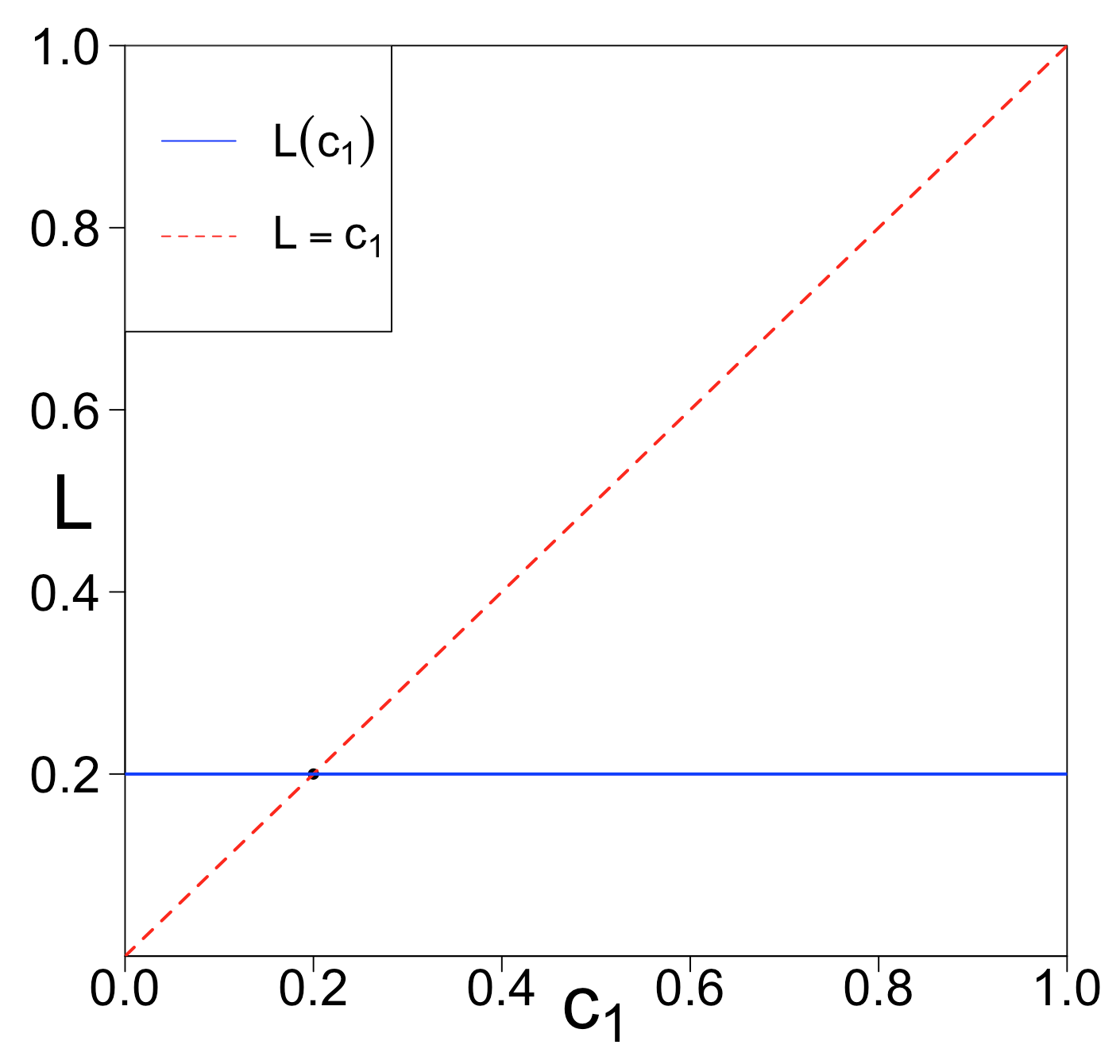}}
	\subfigure[Scenario 2]{\label{fig:L2}\includegraphics[width=50mm]{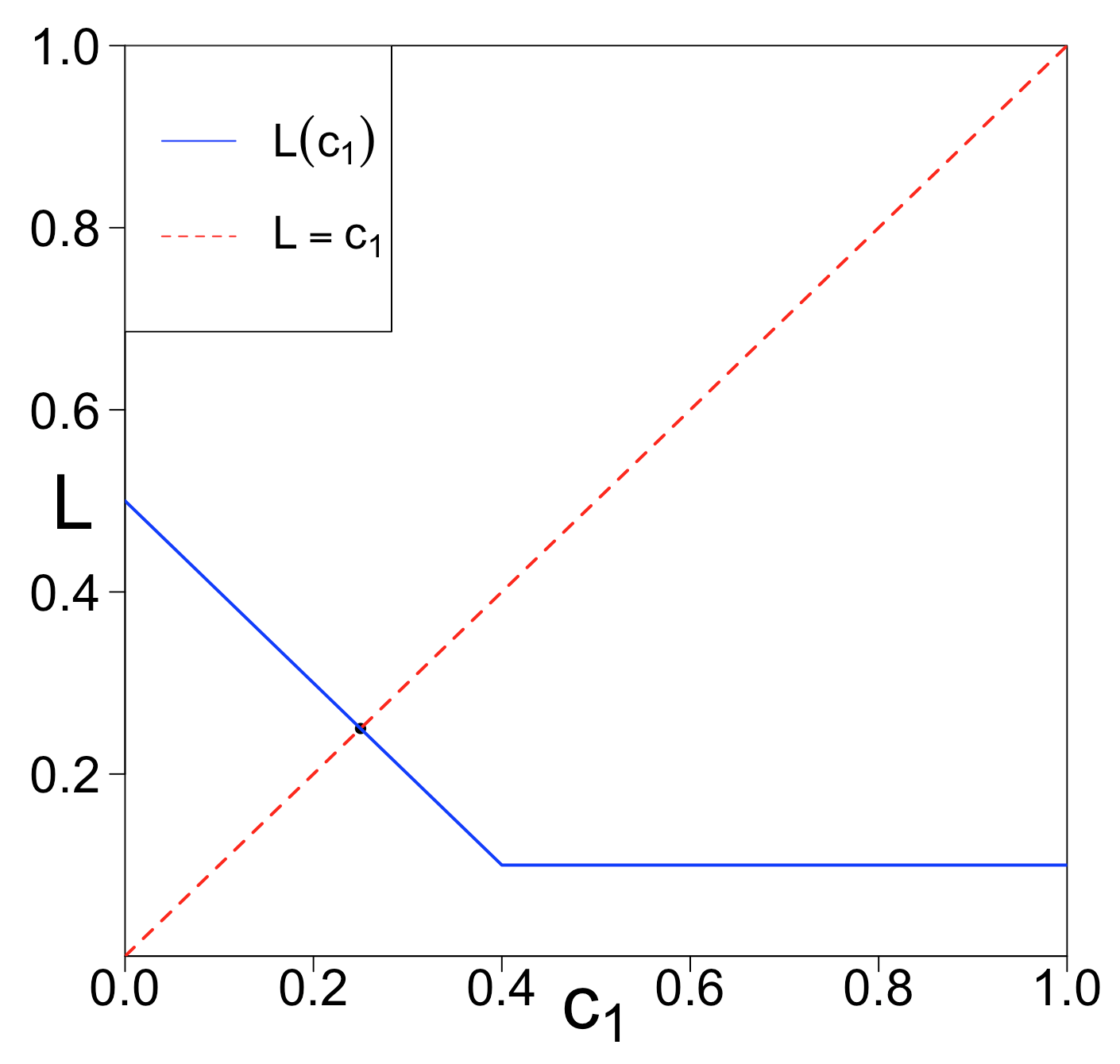}}
	\caption{Two possible scenarios of piecewise linear functions for the upper bound $U$ (top two figures), and lower bound $L$ (lower two figures)}
	\label{fig:piecewise}
\end{figure}

Furthermore, $U_{}$ is a piecewise linear function of $c_{1}$, with slope either 0 or 1. When $c_{1}=1$, we have $\min\{U_{1}(c_{1}),U_{3}(c_{1})\}<\min\{U_{2}(c_{1}),U_{4}(c_{1})\}$. Thus, only two scenarios would occur, as shown in Figures \ref{fig:U1}--\ref{fig:U2}. The curve of  $U_{}(c_{1})$ is either a horizontal line (Scenario 1) or a broken line of two segments with slopes 1 and 0, respectively (Scenario 2). By Criterion \ref{crit: non-strong}, individual surrogate paradox can be excluded if $U_{}(c_{1})\leq c_{1}$. That is, when the red line intersects with or is above the blue line in Figures \ref{fig:U1}--\ref{fig:U2}, individual surrogate paradox can be excluded. 

As for lower bound, $L_{1}$, $L_{3}$ are simple bounds and $L$ is non-trivial when $L_{{2}}\geq \max(L_{{1}},L_{{3}})$ or $L_{{4}}\geq \max(L_{{1}},L_{{3}})$. This is equivalent to one of the two sets of the following inequalities holds:

$$\left\{
\begin{array}{ll} 
P(Y=1,S=0|T=0)<P(Y=1,S=0|T=1),\\ 
P(Y=1,S=1|T=1)<P(Y=1,S=1|T=0);
\end{array}
\right.$$

$$\left\{
\begin{array}{ll} 
P(Y=0,S=1|T=1)<P(Y=0,S=1|T=0),\\ 
P(Y=0,S=0|T=0)<P(Y=0,S=0|T=1).
\end{array}
\right.$$

Similar to the case for the upper bound, the lower bound $L$ is also a piecewise linear function of $c_1$, with slope either 0 or $-1$. And only two scenarios would occur which are depicted in Figures \ref{fig:L1}--\ref{fig:L2}. By Criterion \ref{crit: non-strong}, individual surrogate paradox is guaranteed to be present if $L_{}(c_{1})> c_{1}$. That is, when the red line is below the blue line in Figures \ref{fig:L1}--\ref{fig:L2}, the presence of individual surrogate paradox can be concluded from the observed data. 

Although the bounds in Proposition \ref{prop: non-strong} are obtained for binary surrogate and outcome, similar methods can be easily extended to the case where surrogate and outcome are multi-level variables. Once we obtain the bounds, the criterion to exclude or conclude the presence of individual surrogate paradox can be obtained in the same fashion.  

The bounds obtained in Proposition \ref{prop: non-strong} can be wide since no additional information is incorporated. Conditions could be further imposed to tighten the bound and hence increase power for the corresponding criteria. For illustration, we impose an additional assumption that is motivated by the causal necessity condition. Recall that \citet{frangakis2002principal} proposed the causal necessity condition as follows.

\begin{defn}[\textbf{Causal Necessity}]
	\vspace{8pt}
	 Assume $T$ does not affect $Y$ if $T$ does not affect $S$, i.e., if $S(0) = S(1)$, then $Y(0) = Y(1)$.
\end{defn}

\vspace{8pt}
\noindent Causal Necessity states that the treatment $T$ can affect primary outcome $Y$ only through surrogate $S$. But here we generalize this idea. We set $0\leq c_3\leq1$ as the proportion of the individuals that do not conform causal necessity. That is, 

\begin{condition}\label{cond: neces}
Assume  the proportion of individuals that have the same surrogate under treatment and control but different primary outcome is no greater than $c_3$, i.e., $P(S_1=S_0,Y_{1S_1}\neq Y_{0S_0})\leq c_3$.
\end{condition}
The causal necessity condition is a special case of the Condition \ref{cond: neces} when $c_3=0$. While causal necessity excludes any one from having a non-null treatment effect on primary outcome if there is no treatment effect on surrogate endpoint, Condition \ref{cond: neces} allows a small proportion of violation. We update our bounds under Condition \ref{cond: neces} to gain more power.

%Recall the 64 potential outcome types in the strong surrogate scenario. $P(S_1=S_0,Y_{1S_1}\neq Y_{0S_0})$=$q_{2,0}+q_{3,0}+q_{6,0}+q_{7,0}+q_{8,0}+q_{9,0}+q_{12,0}+q_{13,0}+q_{1,3}+q_{3,3}+q_{6,3}+q_{9,3}+q_{11,3}+q_{12,3}+q_{14,3}$. Add this restriction into the linear programming project, we see that upper bound indeed shrinks but lower bound remains the same.

\begin{prop}\label{prop: causal necess}
	Given $P(Y,S|T)$ and assume Condition \ref{cond: neces}  holds, if $\mathrm{HR}(T\rightarrow S)\leq c_1$ and $P(S_1=S_0,Y_{1S_1}\neq Y_{0S_0})\leq c_3$, then $L\leq \mathrm{HR}(T\rightarrow Y)\leq \tilde U_{},$
	where $L=\max(L_{1},~L_{2},~L_{3},~L_{4})$, $\tilde U=\min(\tilde U_1,\ldots,\tilde U_{15})$, and the formula of $\tilde U_1,\ldots, \tilde U_{15}$ is given in the Supplementary material.
\end{prop}

Similar as in Proposition \ref{prop: non-strong}, both $L$ and $\tilde U$ only involve observed data distribution, thus, they can be calculated. It can also be shown that $\tilde U\leq U$, hence, the upper bound is tighter under additional Assumption \ref{cond: neces}. Interestingly, the incorporation of Condition \ref{cond: neces} only shrink the upper bound with the lower bound unchanged.

According to Proposition \ref{prop: causal necess}, we have the following criterion to exclude or conclude the existence of individual surrogate paradox.

\begin{crit} If Condition \ref{cond: neces} holds, then
	(1) the individual surrogate paradox can be excluded if observed distribution $P(Y,S|T)$ satisfies $\tilde U_{}=\min(\tilde U_{1},\ldots, ~\tilde U_{15})\leq c_{1}$;
	(2) the individual surrogate paradox is guaranteed to manifest if $L=\max(L_{1},~L_{2},~L_{3},~L_{4})> c_1.$
\end{crit}

\section{Data analysis}\label{sec: data}

%We now apply the methods developed in Proposition \ref{prop: non-strong} and \ref{prop: causal necess} to investigate the the glycated hemoglobin level as a valid surrogate to evaluate the effect of intensive glycemia drugs on the diabetic retinopathy. We will use the data in The Action to Control Cardiovascular Risk in Diabetes (ACCORD). 

Diabetic retinopathy is caused by damage to the blood vessels in the light sensitive tissue at the back of the eye (retina). It is a leading cause of vision loss among people with diabetes and the major cause of vision damage and blindness among working-age adults. The ACCORD study enrolled 10,251 participants with type 2 diabetes who were at high risk for cardiovascular disease to randomly receive either intensive or standard treatment for glycemia (target glycated hemoglobin level, $<$ 6.0\% or 7.0 to 7.9\%, respectively, The ACCORD Study Group \nocite{doi:10.1056/NEJMoa1001288} and ACCORD Eye Study Group, 2010). The ACCORD Eye study aimed at determining the effect of the intensive glycemia on the risk of development or progression of diabetic retinopathy. Among all the ACCORD participants, 2856 were eligible for the ACCORD Eye study. Since the inclusion criteria did not affect the intervention, we still consider the ACCORD Eye study as a randomized trial.
%\begin{figure}[!h]
%	\centering
%	\begin{minipage}{.5\textwidth}
%		\centering
%		\includegraphics[width=1\linewidth]{Figure3-1.png}
%		\captionof{figure}{$U(c_1)$ when $c_3$=0}
%	\end{minipage}%
%	\begin{minipage}{.5\textwidth}
%		\centering
%		\captionof{figure}{$U(c_1)$ when $c_3$=0.02}
%	\end{minipage}
	
%	\begin{minipage}{.5\textwidth}
%		\centering
%		\includegraphics[width=1\linewidth]{Figure3-3.png}
%		\captionof{figure}{$U(c_1)$ when $c_3$=0.05}
%	\end{minipage}%
%	\begin{minipage}{.5\textwidth}
%		\centering
%		\includegraphics[width=1\linewidth]{Figure3-4.png}
%		\captionof{figure}{$U(c_1)$ when $c_3$=0.1}
%	\end{minipage}
%	
%\end{figure}

Let $Y=0$ if the progression of diabetic retinopathy was observed, and $Y=1$ otherwise. Let $S$ denote the glycated hemoglobin level. We set $S=1$ if it was smaller than 6\% at one year follow-up time and $S=0$ otherwise. Moreover, let $T=1$ if the patient gets intensive treatment and $T=0$ otherwise. 

We first calculate the bound of $\mathrm{HR}(T\rightarrow Y)$ without any additional assumptions. By Proposition \ref{prop: non-strong}, the bound of $\mathrm{HR}(T\rightarrow Y)$ is [0.033, 0.105] and the bound does not change with $c_1$. Using bootstrap, we obtain the 95\% uncertainty region (confidence interval for bounds, see more in \citet{richardson2014nonparametric}) of $\mathrm{HR}(T\rightarrow Y)$ as [0.011, 0.119]. That is, under 95\% confidence level, for any $c_1 > 0.119$, if the proportion of individuals that have glycated hemoglobin level increased to $>6\%$  in one year follow up due to intensive treatment is less than $c_1$, then, no more than $c_1$ proportion of individuals will get progression of diabetic retinopathy due to the intensive treatment.  On the other hand, under 95\% confidence level, for any $c_1 < 0.011$, even if the proportion of individuals that have glycated hemoglobin level increased to $>6\%$  in one year follow up due to intensive treatment is less than $c_1$, more than $c_1$ proportion  of the population will get progression of diabetic retinopathy due to the intensive treatment. 

\begin{figure}
	\centering    
	\subfigure[$c_3$=0]{\label{fig:c3=0}\includegraphics[width=50mm]{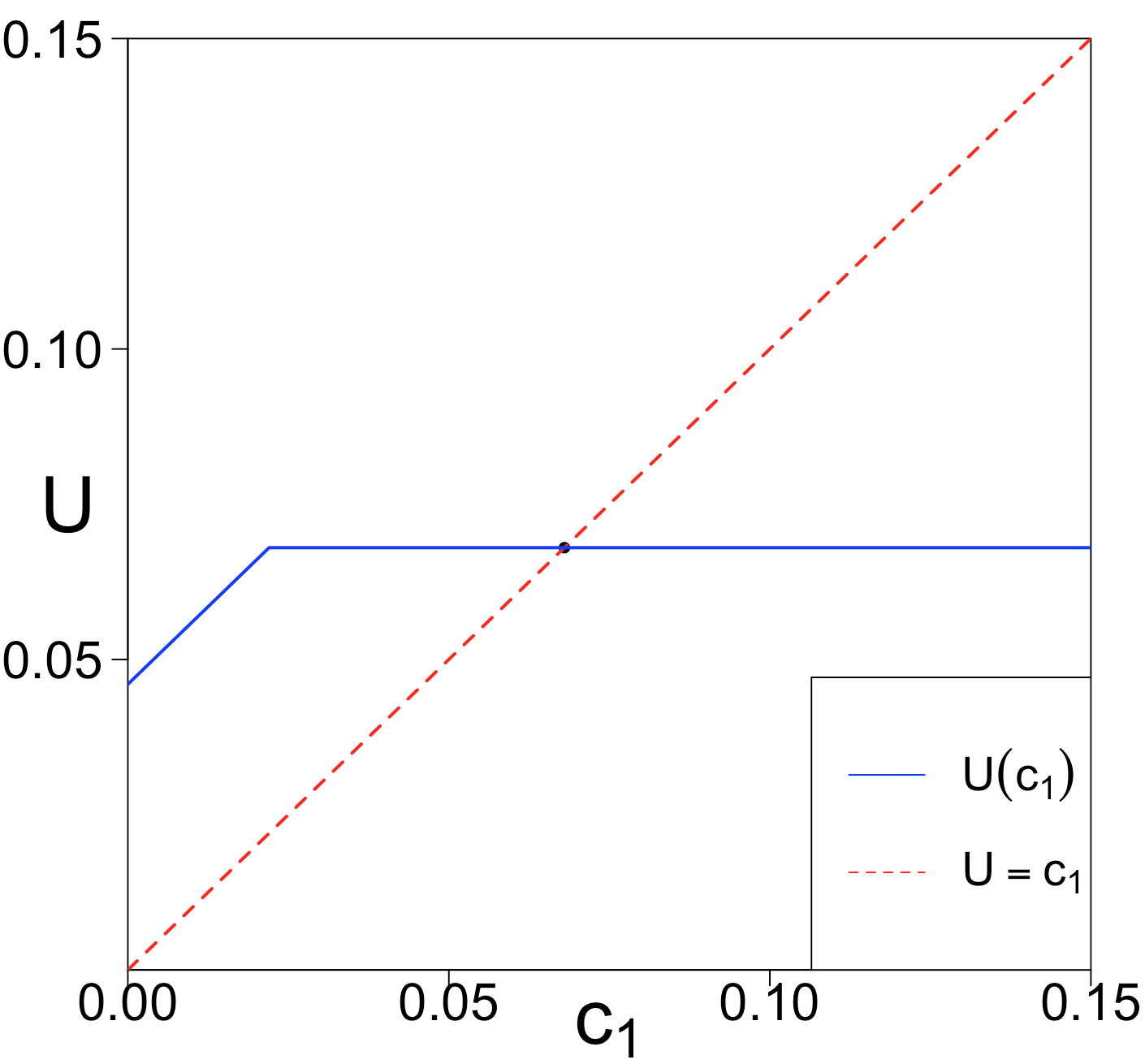}}
	\subfigure[$c_3$=0.02]{\label{fig:c3=0.02}\includegraphics[width=50mm]{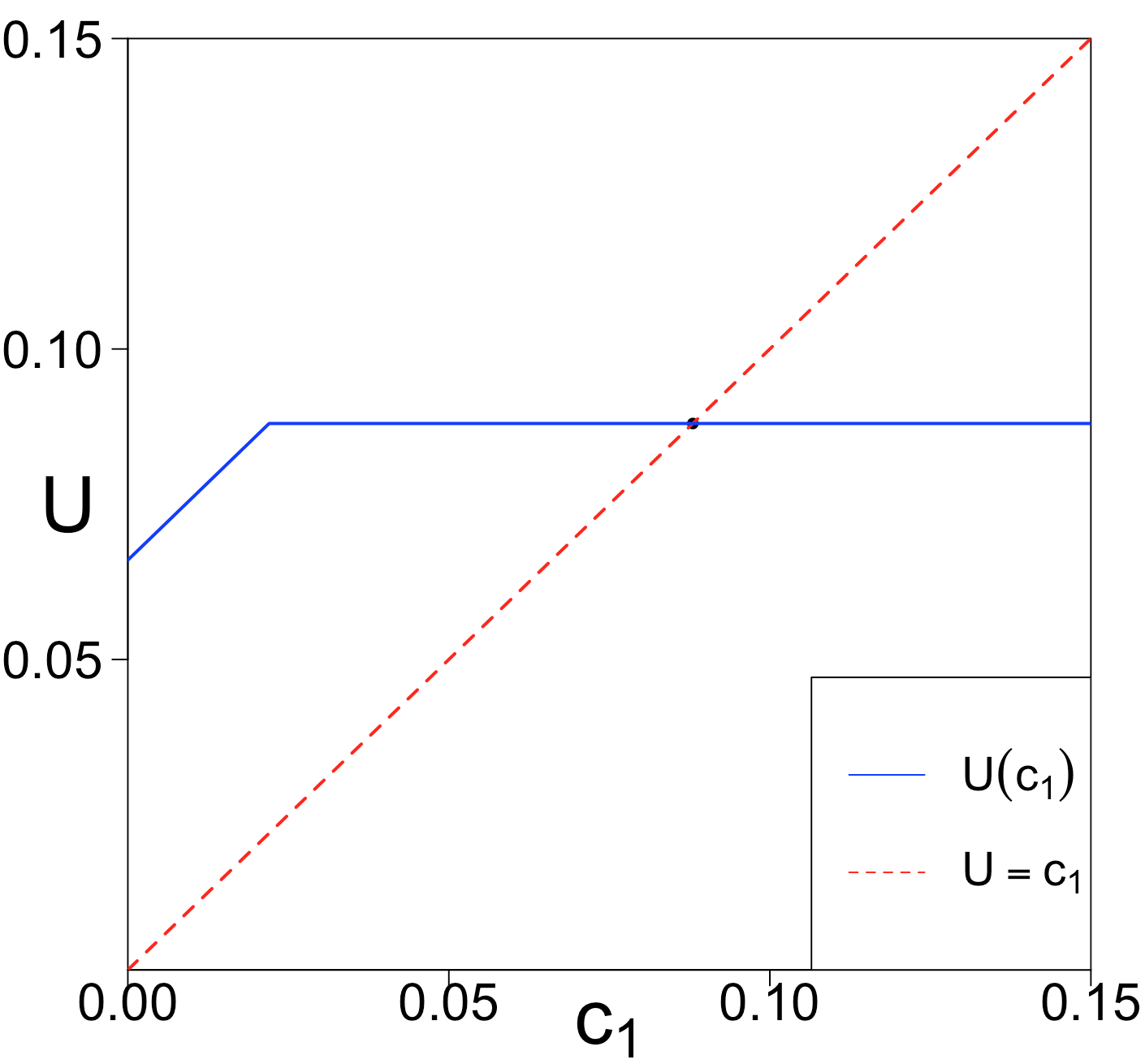}}
	
	\subfigure[$c_3$=0.05]{\label{fig:c3=0.05}\includegraphics[width=50mm]{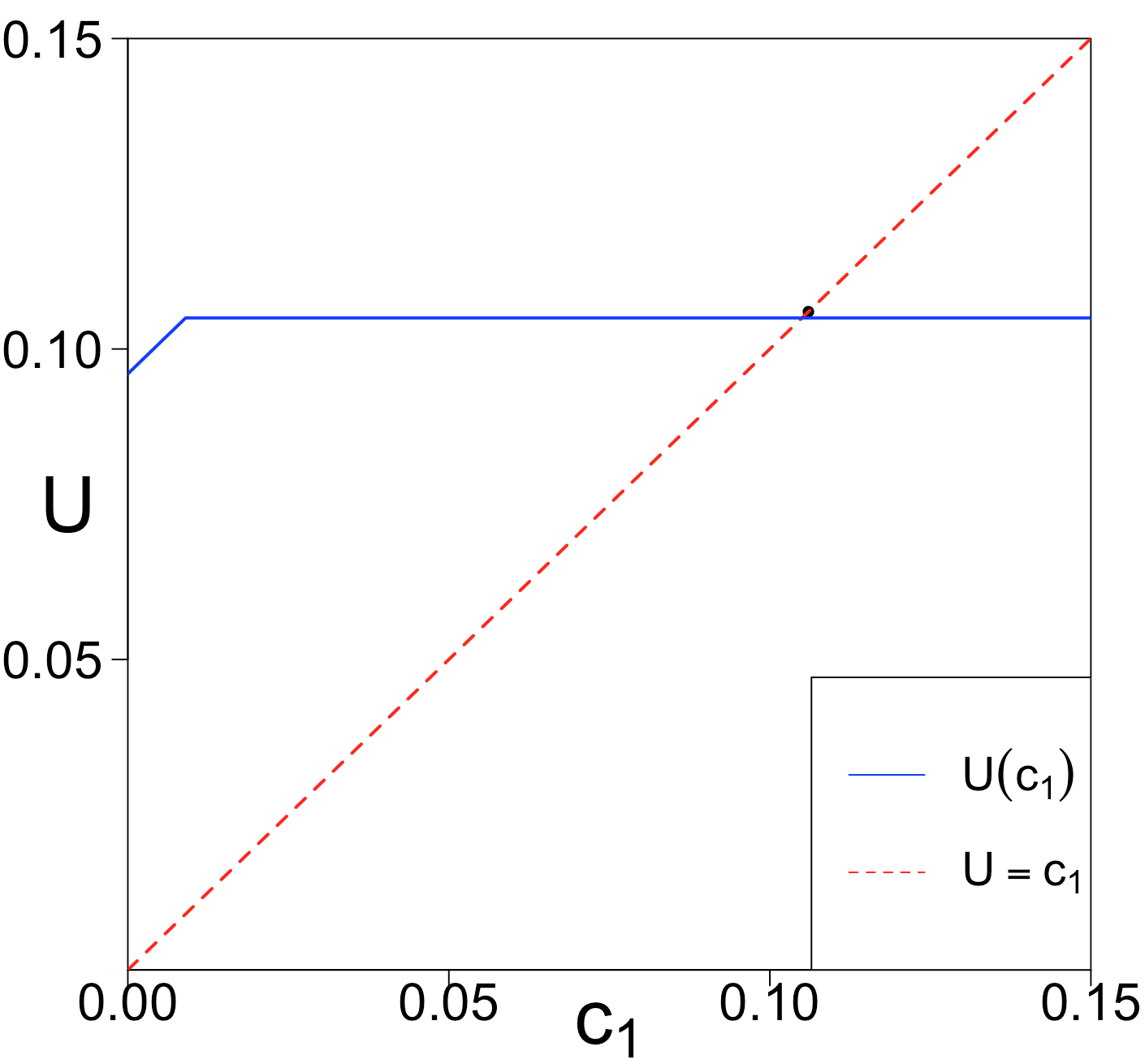}}
	\subfigure[$c_3$=0.1]{\label{fig:c3=0.1}\includegraphics[width=50mm]{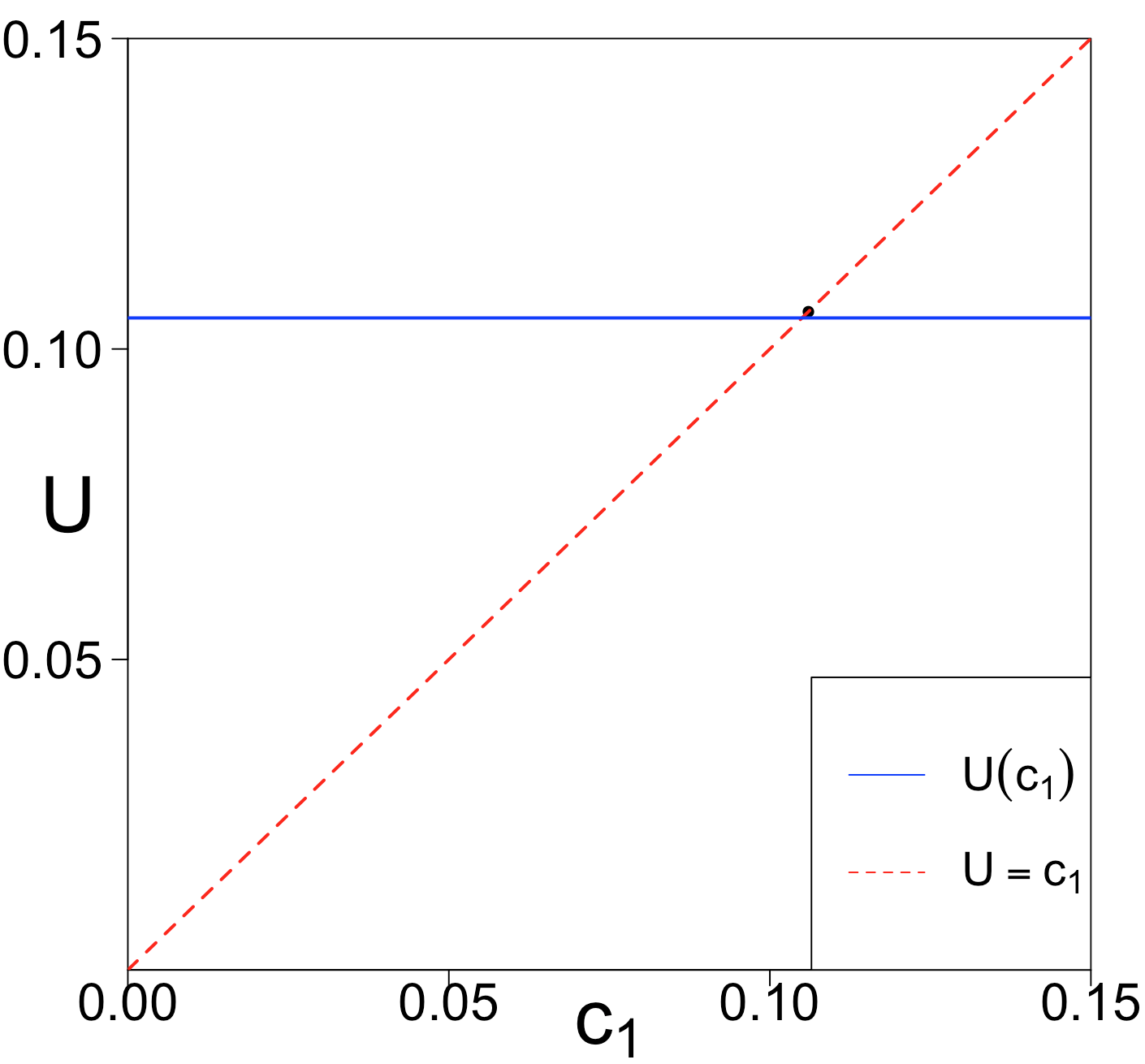}}
	\caption{Curves of $U_{}(c_1)$ with different values of $c_3$}
	\label{fig:data_analysis}
\end{figure}

Under Condition \ref{cond: neces}, we calculate the upper bound of $\mathrm{HR}(T\rightarrow Y)$ for different values of $c_3$ since lower bound is unchanged under this additional condition. As shown in Figure \ref{fig:c3=0}, when $c_3=0$, i.e., causal necessity holds, by Proposition \ref{prop: causal necess}, we have $\tilde U=0.046+c_1$ for $c_1\leq 0.022$, and $\tilde U=0.068$ otherwise. The 95\% uncertainty region is [0.025$+c_1$, 0.067$+c_1$] for $c_1\leq 0.022$ and [0.047, 0.089] otherwise. Thus, when $c_1>0.089$, under 95\% confidence level, the individual surrogate paradox can be excluded, i.e., there is no more individual get progression of diabetic retinopathy due to the intensive treatment than those who have a harmful treatment effect on the glycated hemoglobin level. A comparison with the bounds without the causal necessity reveals that the incorporation of causal necessity shrinks the upper bound significantly. As shown in Figure \ref{fig:c3=0.02}, when $c_3=0.02$, the upper bound inflates to  $\tilde U=0.066+c_1$ for $c_1\leq 0.022$, and $\tilde U=0.088$ otherwise. The 95\% uncertainty region is [0.045$+c_1$, 0.089$+c_1$] and [0.067, 0.111], respectively. Hence, relaxing the causal necessity condition makes it harder to exclude the individual surrogate paradox. As shown in Figure \ref{fig:c3=0.05}, when $c_3=0.05$, we have $\tilde U=0.096+c_1$ when $c_1\leq 0.009$, and $\tilde U=0.105$ otherwise. The 95\% uncertainty region is [0.079$+c_1$, 0.113$+c_1$] and [0.088, 0.122]. Finally, when $c_3\geq 0.059$, the generalized causal necessity assumption can no longer further shrink the bounds as compared to those without the assumption. Hence, the bounds are [0.033, 0.105] with 95\% uncertainty region [0.011, 0.119] (see Figure \ref{fig:c3=0.1}), which is the same as the result we obtain from Proposition \ref{prop: non-strong}. %Recall that from Proposition \ref{prop: causal necess}, lower bound is independent of $c_3$, so the paradox manifests in all the scenarios mentioned above with 95\% confidence level when $c_1<0.011$.

%We drew four figures to make it easier to find the relationship between $U$ and $c_1$ when $c_3$ varies. From the figures we know that if we can estimate the upper bound of $c_3$ in an experiment, then we can use $c_3$ to exclude paradox more easily. 

%\vspace{-8mm}
\section{Discussion}\label{sec: discuss}
Compared to the surrogate paradoxes based on ACE and DCE, the individual surrogate paradox we proposed focuses more on the individuals. In this paper, we propose criterion to exclude the individual surrogate paradox based on sharp bounds of $\mathrm{HR}(T\rightarrow Y)$. However, without any assumptions, the bounds may be wide since it accommodates all the possible scenarios and hence the criterion may not have enough power to exclude the paradox. To increase power of our criteria, we impose a generalized causal necessity assumption. Such assumption facilitates the exclusion of individual surrogate paradox but does not change the conclusion of the existence of such paradox. Other assumptions may be imposed to further influence both the lower and upp	er bounds. We leave that as future research topic. We also leave the exclusion of non-strong surrogate in the continuous case as a future research topic.

\bibliographystyle{apa}
%\bibliography{bibfile1}
\bibliography{surrogate}

\newpage
\appendix

\section*{Supplementary Materials}
\subsection*{Additional Counterexample for Prentice Criterion} 

The following example illustrates that the Prentice criterion cannot exclude the existence of individual surrogate paradox in the continuous case.

\begin{example}\label{eg: prentice}
	When surrogate $S$ and outcome $Y$ are both continuous, assume the following model holds:
	\begin{equation}\label{model_prentice_con}
	\left\{
	\begin{array}{ll}
	Y=\alpha_{0}+\alpha_{1}T+\alpha_{2}S+\alpha_{3}W+\epsilon_{1} ,  \\
	S=\beta_{0}+\beta_{1}T+\beta_{2}W+\epsilon_{2} ,  \\
	\epsilon_{1}\sim N(0,1),~~\epsilon_{2}\sim N(0,1),~~ W\sim N(0,1),~~ (\epsilon_{1},\epsilon_{2})\bot (T,S,W),~~ \epsilon_{1}\bot\epsilon_{2} .
	\end{array}
	\right.
	\end{equation}
	If $\beta_{1}>0$, $\alpha_{2}>0$, $\alpha_{1}(1+\beta_{2}^{2})-\alpha_{3}\beta_{1}\beta_{2}=0$ and $\alpha_{1}+\alpha_{2}\beta_{1}<0$, then, the individual surrogate paradox manifests while Prentice criterion is satisfied.
	\label{example2}
\end{example}

\noindent There are many choices of parameter values that satisfy the conditions in Example \ref{eg: prentice}.  For example, we set the parameter value as $\alpha_{1}=-2,~\alpha_{2}=1,~\alpha_{3}=-4,~\beta_{1}=1,~\beta_{2}=1.$ We first show that the Prentice criterion holds. When $\beta_{1}>0$, $T$ does no harm to $S$, i.e., $\mathrm{HR}(T\rightarrow S)=0$. When $\alpha_{2}>0$, $S$ does no harm to $Y$, i.e., $\mathrm{HR}_{s_0,s_1}(S\rightarrow Y|T=t)>0$. We prove at the end of Supplementary Materials that given $(T,S)$, the distribution of $Y$ is
$$Y|(T,S)\sim N\left\{\mu(T,S), ~\frac{1+\beta_{2}^{2}}{2(1+\beta_{2}^{2}+\alpha_{3}^{2})}\right\},$$
where $\mu(t,s)=\big[\big\{(1+\beta_{2}^{2})\alpha_{0}-\alpha_{3}\beta_{0}\beta_{2}\big\}
+\big\{\alpha_{1}(1+\beta_{2}^{2})-\alpha_{3}\beta_{1}\beta_{2}\big\}t
+\big\{\alpha_{2}(1+\beta_{2}^{2})+\alpha_{3}\beta_{2}\big\}s\big]/(1+\beta_{2}^{2})$. When $\alpha_{1}(1+\beta_{2}^{2})-\alpha_{3}\beta_{1}\beta_{2}=0$, we have $Y\bot S|T$, which means Prentice criterion holds. On the other hand, model \eqref{model_prentice_con} indicates that
$$Y=(\alpha_{0}+\alpha_{2}\beta_{0})+(\alpha_{1}+\alpha_{2}\beta_{1})T+(\alpha_{3}+\alpha_{2}\beta_{2})W
+(\epsilon_{1}+\alpha_{2}\epsilon_{2}),$$
\noindent hence, $Y_{T=1}-Y_{T=0}=\alpha_{1}+\alpha_{2}\beta_{1}.$ When $\alpha_{1}+\alpha_{2}\beta_{1}<0$, there exist individuals who has a negative treatment effect on primary outcome $Y$, i.e., $\mathrm{HR}(T\rightarrow Y)>0$. This implies when $S$ and $Y$ are continuous, Prentice criterion cannot avoid the individual surrogate paradox. 

\subsection*{Additional counterexample for principal surrogate criterion}

\begin{example}
	When surrogate $S$ and outcome $Y$ are both continuous, assume the following model holds:
	\begin{equation}\label{model_principal_con}
	\left\{
	\begin{array}{ll}
	Y=\alpha_{0}+\alpha_{1}S+\alpha_{2}W+(2T-1)\epsilon_{1} ,  \\
	S=\beta_{0}+\beta_{1}T+\beta_{2}W+\epsilon_{2} ,  \\
	\epsilon_{1}\sim N(0,1),~~\epsilon_{2}\sim N(0,1),~~ W\sim N(0,1),~~ (\epsilon_{1},\epsilon_{2})\bot (T,S,W),~~ \epsilon_{1}\bot\epsilon_{2} .
	\end{array}
	\right.
	\end{equation}
	If $\beta_{1}>0$, $\alpha_{1}>0$,  then, principle surrogate criterion holds whereas individual surrogate paradox manifests.
	\label{example4}
\end{example}

The difference between this model and model \eqref{model_prentice_con} is that in this model, the effect of $T$ on $Y$ is on a error variable $\epsilon_{1}$, but the effect of $T$ on $Y$ in model \eqref{model_prentice_con} is on the interception. When $\beta_{1}>0$, $T$ does no harm to $S$ for any individual, i.e., $\mathrm{HR}(T\rightarrow S)=0$. When $\alpha_{1}>0$, $S$ does no harm to $Y$ for any individual, i.e., $\mathrm{HR}_{s_0,s_1}(S\rightarrow Y|T=t)=0$. Note
\begin{eqnarray*}
	&&P(Y_{T}|S_{0}=S_{1}=s) \\
	 &=& \int P(Y_{T}|S_{0}=S_{1}=s,W=w)P(W=w|S_{0}=S_{1}=s) du\\
	&=& \int P(\alpha_{0}+\alpha_{1}s+\alpha_{2}u+(2T-1)\epsilon_{1}|S_{0}=S_{1}=s,W=w)P(W=w|S_{0}=S_{1}=s) du,
\end{eqnarray*}
we also have
$$\{\alpha_{0}+\alpha_{1}s+\alpha_{2}u+(2T-1)\epsilon_{1}\}|(S_{0}=S_{1}=s,W=w)\sim N(\alpha_{0}+\alpha_{1}s+\alpha_{2}u, 1),$$
for $t=0,1$. Hence, we have $P(Y_{T=0}|S_{0}=S_{1}=s)=P(Y_{T=1}|S_{0}=S_{1}=s)$, which indicates that under model \eqref{model_principal_con}, $S$ is a principle surrogate. 

On the other hand, according to model \eqref{model_principal_con}, we have:
$$Y=(\alpha_{0}+\alpha_{1}\beta_{0})+\alpha_{1}\beta_{1}T+(\alpha_{2}+\alpha_{1}\beta_{2})W+(2T-1)\epsilon_{1}+\alpha_{1}\epsilon_{2}.$$
Hence $\mathrm{HR}(T\rightarrow Y)=P(Y_{T=0}-Y_{T=1}>0)=P(-\alpha_{1}\beta_{1}-2\epsilon_{1}>0)=\Phi(-\alpha_{1}\beta_{1}/2),$
where $\Phi(\cdot)$ is the function of standard normal distribution. Set $\alpha_{1}=\beta_{1}=1$, we obtain $T$ does no harm to $S$ and $S$ does no harm to $Y$, but $\mathrm{HR}(T\rightarrow Y)=\Phi(-1/2)>0$. Hence, the individual surrogate paradox manifests. %Therefore, when $S$ and $Y$ are all continuous, principle surrogate cannot avoid the individual surrogate paradox. 

\subsection*{Counterexample for strong surrogate criterion}

%Although the strong surrogate criterion can exclude the individual surrogate paradox for binary surrogate, it cannot exclude it for non-binary surrogate. %The following example provides a counterexample when the strong surrogate criterion is satisfied for a surrogate that has three possible values, but the individual surrogate paradox still manifests. 

\begin{example}\label{ex:strong}
	When $S$ takes value of 0, 1, 2 and $Y$ is binary. We assume $S$ is a strong surrogate in this case. Then, we have $3^2\times 2^3 = 72$ different possible values for the vector $(S_0, S_1, Y_0, Y_1, Y_2)$.  Let $q_{i,j}\geq0$ denote the proportion of getting each possible value of $(S_0, S_1, Y_0, Y_1, Y_2)$ in the whole population for $i=0,\ldots,8$ and $j=0,\ldots,7$ (shown in Table \ref{tb: counterexample_q} in the Supplementary Materials). We have $\mathrm{HR}(T\rightarrow S)=\sum_{i=3,6,7}\sum_{j=0}^{8}q_{i,j}$, $\mathrm{HR}(S\rightarrow Y)=\max(c_{21}, c_{22}, c_{23})$, where $c_{21}=P(Y_{S=0}>Y_{S=1})=\sum_{i=0}^{8}\sum_{j=4,5}q_{i,j}$, $c_{22}=P(Y_{S=0}>Y_{S=2})=\sum_{i=0}^{8}\sum_{j=4,6}q_{i,j}$, $c_{23}=P(Y_{S=1}>Y_{S=2})=\sum_{i=0}^{8}\sum_{j=2,6}q_{i,j}$.
	And $\mathrm{HR}(T\rightarrow S)=P(Y_{T=0}=1, Y_{T=1}=0)=q_{6,1}+q_{7,1}+q_{3,2}+q_{5,2}+q_{3,3}+q_{6,3}+q_{1,4}+q_{2,4}+q_{1,5}+q_{7,5}+q_{2,6}+q_{5,6} $. Let $q_{5,2}=q_{1,4}=q_{2,4}=q_{1,5}=q_{2,6}=q_{5,6}=1/6$, and $q_{i,j}=0$ otherwise. Then, we have $\mathrm{HR}(T\rightarrow S)=0$, $\mathrm{HR}(S\rightarrow Y)=2/3$, but $\mathrm{HR}(T\rightarrow Y)=1>\mathrm{HR}(T\rightarrow S)+\mathrm{HR}(S\rightarrow Y).$ Thus, individual surrogate manifests.
\end{example}

\subsection*{Counterexample for Wu and VanderWeele Criteria}
We use example \ref{example1} as the counterexample for Wu Criterion. From Table \ref{tb: eg_prentice}, we have $f(1,1)=P(Y=1|S=1,T=1)=0.75$, $f(1,0)=P(Y=1|S=1,T=0)=0.75$, $f(0,1)=P(Y=1|S=0,T=1)=0.5$, and $f(0,0)=P(Y=1|S=0,T=0)=0.5$. Hence, we have $f(1,1)\geq f(0,1)$, $f(1,0)\geq f(0,0)$, $f(0,1)=f(0,0)$, $f(1,1)=f(1,0)$ satisfying the condition of \citet{wu2011sufficient}. However, we have $\mathrm{HR(T\rightarrow Y)}=0.235>0$. Thus, individual surrogate paradox manifests.

To verify VanderWeele criterion cannot exclude the individual surrogate paradox, we consider a slightly modified version of example \ref{example1}. When both $S$ and $Y$ are binary, we have $\mathrm{HR}(T\rightarrow Y)=P(Y_{T=0}=1,Y_{T=1}=0)=q_{4,3}+q_{12,0}+q_{12,1}+q_{12,3}+q_{13,0}.$
Assume $W$ is binary with $ P(W=1)=P(W=0)=0.5$ and the joint distribution of $(S_0,S_1,Y_{00},Y_{01}, Y_{10},Y_{11})|W=w$ the same as the probabilities in given in Table \ref{tb: eg_prentice} for $w=0,1$. For example, assume $P(Y_{00}=1, S_0=0|W=w)=P(Y_{00}=1, S_0=0)= 0.3$. Then, we have
$$\left\{
\begin{array}{ll}
P(Y=1|T=0,S=0,W=w)=\dfrac{P(Y_{00}=1,S_{0}=0|W=w)}{P(S_{0}=0|W=w)}
=0.5   ,  \\
P(Y=1|T=1,S=0,W=w)=\dfrac{P(Y_{10}=1,S_{1}=0|W=w)}{P(S_{1}=0|W=w)}
=0.5   ,  \\
P(Y=1|T=0,S=1,W=w)=\dfrac{P(Y_{01}=1,S_{0}=1|W=w)}{P(S_{0}=1|W=w)}
=0.75   ,  \\
P(Y=1|T=1,S=1,W=w)=\dfrac{P(Y_{11}=1,S_{1}=1|W=w)}{P(S_{1}=1|W=w)}
=0.75   .
\end{array}
\right.$$
Therefore, condition (a) is satisfied. Additionally, we have $P(S=1|T=0,w)=0.4\leq 0.8=P(S=1|T=1,w)$ satisfying condition (b).
Since $\mathrm{HR}(T\rightarrow Y)=P(Y_{T=0}=1,Y_{T=1}=0)=q_{4,3}+q_{12,0}+q_{12,1}+q_{12,3}+q_{13,0}=0.235>0$, $T$ does harm to $Y$ for 23.5\% of the individuals. %Individual surrogate paradox happens in this example.

\subsection*{The proof of Proposition 1}
 When $P(Y,S|T)$ is known, we have the following restrictions on the distributional parameters of the potential outcomes:
\begin{equation}\label{constrain1}
\left\{
\begin{array}{ll}
P(Y=0, S=0|T=0)=$$\sum_{i=0}^{7} (q_{i,0}+q_{i,1})  ,  \\
P(Y=0, S=1|T=0)=$$\sum_{i=0,1,2,3,8,9,10,11} (q_{i,2}+q_{i,3})  ,  \\
P(Y=1, S=0|T=0)=$$\sum_{i=8}^{15} (q_{i,0}+q_{i,1})  ,  \\
P(Y=0, S=0|T=1)=$$\sum_{i=0,1,4,5,8,9,12,13} (q_{i,0}+q_{i,2})  ,  \\
P(Y=0, S=1|T=1)=$$\sum_{i=0,2,4,6,8,10,12,14} (q_{i,1}+q_{i,3})  ,  \\
P(Y=1, S=0|T=1)=$$\sum_{i=2,3,6,7,10,11,14,15} (q_{i,0}+q_{i,2})  ,  \\
1=\sum_{i}(q_{i,0}+q_{i,1}+q_{i,2}+q_{i,3})  ,  
\end{array}
\right.
\end{equation}

Since we have $$\mathrm{HR}(T\rightarrow S)=P(S_{T=0}=1,S_{T=1}=0)\leq c_{1},$$
$$\mathrm{HR}(S\rightarrow Y|T=0)=P(Y_{00}=1,Y_{01}=0)\leq c_{2},$$
$$\mathrm{HR}(S\rightarrow Y|T=1)=P(Y_{10}=1,Y_{11}=0)\leq c_{2}.$$
Therefore, we have the following restrictions:
\begin{equation}\label{constrain2}
\left\{
\begin{array}{ll}
$$\sum_{i=0}^{15} q_{i,2}+\alpha=c_{1}  ,  \\
$$\sum_{j=0,1,2,3} (q_{8,j}+q_{9,j}+q_{10,j}+q_{11,j})+\beta=c_{2}  ,  \\
$$\sum_{j=0,1,2,3} (q_{2,j}+q_{6,j}+q_{10,j}+q_{14,j})+\gamma=c_{2}  ,  \\
\alpha,\beta,\gamma\geq 0  ,  \\
q_{i,j}\geq0  , \text{ for any } i,j.
\end{array}
\right.
\end{equation}

Our aim is to determine the bound of
$\mathrm{HR}(T\rightarrow Y)=q_{8,0}+q_{9,0}+q_{12,0}+q_{13,0}+q_{8,1}+q_{10,1}+q_{12,1}+q_{14,1}
+q_{4,2}+q_{5,2}+q_{12,2}+q_{13,2}+q_{4,3}+q_{6,3}+q_{12,3}+q_{14,3}.$ \label{defHR}
Let $\overrightarrow{q}=(q_{0,0},q_{1,0},\ldots,q_{15,0},q_{1,0},\ldots,q_{15,3},\\\alpha,\beta,\gamma)'$, $b=(P(Y=0,S=0|T=0), P(Y=0,S=1|T=0), P(Y=1,S=0|T=0), P(Y=0,S=0|T=1), P(Y=0,S=1|T=1), P(Y=1,S=0|T=1), 1, c_{1}, c_{2},c_{2})'$, 
$c=(0,0,0,0,0,0,0,0,1,1,0,0,1,1,0,0,0,0,0,0,0,0,0,0,1,0,1,0,1,0,1,0,0,0,0,0,1,1,0,0,0,0,\\0,0,1,1,0,0,0,0,0,0,1,0,1,0,0,0,0,0,1,0,1,0,0,0,0)'.$ Denote $A'=(B',C',D',E')$, where
\[
B=\begin{bmatrix}
1&0&0&1&0&0&1&0&0&0\\
1&0&0&1&0&0&1&0&0&0\\
1&0&0&0&0&1&1&0&0&1\\
1&0&0&0&0&1&1&0&0&0\\
1&0&0&1&0&0&1&0&0&0\\
1&0&0&1&0&0&1&0&0&0\\
1&0&0&0&0&1&1&0&0&1\\
1&0&0&0&0&1&1&0&0&0\\
0&0&1&1&0&0&1&0&1&0\\
0&0&1&1&0&0&1&0&1&0\\
0&0&1&0&0&1&1&0&1&1\\
0&0&1&0&0&1&1&0&1&0\\
\end{bmatrix}
,\hspace{1cm}
C=\begin{bmatrix}
0&0&1&1&0&0&1&0&0&0\\
0&0&1&1&0&0&1&0&0&0\\
0&0&1&0&0&1&1&0&0&1\\
0&0&1&0&0&1&1&0&0&0\\
1&0&0&0&1&0&1&0&0&0\\
1&0&0&0&0&0&1&0&0&0\\
1&0&0&0&1&0&1&0&0&1\\
1&0&0&0&0&0&1&0&0&0\\
1&0&0&0&1&0&1&0&0&0\\
1&0&0&0&0&0&1&0&0&0\\
1&0&0&0&1&0&1&0&0&1\\
1&0&0&0&0&0&1&0&0&0\\
\end{bmatrix},
\]
\[
D=\begin{bmatrix}
0&0&1&0&1&0&1&0&1&0\\
0&0&1&0&0&0&1&0&1&0\\
0&0&1&0&1&0&1&0&1&1\\
0&0&1&0&0&0&1&0&1&0\\
0&0&1&0&1&0&1&0&0&0\\
0&0&1&0&0&0&1&0&0&0\\
0&0&1&0&1&0&1&0&0&1\\
0&0&1&0&0&0&1&0&0&0\\
0&1&0&1&0&0&1&1&0&0\\
0&1&0&1&0&0&1&1&0&0\\
0&1&0&0&0&1&1&1&0&1\\
0&1&0&0&0&1&1&1&0&0\\
0&0&0&1&0&0&1&1&0&0\\
0&0&0&1&0&0&1&1&0&0\\
0&0&0&0&0&1&1&1&0&1\\
0&0&0&0&0&1&1&1&0&0\\
0&1&0&1&0&0&1&1&1&0\\
0&1&0&1&0&0&1&1&1&0\\
0&1&0&0&0&1&1&1&1&1\\
0&1&0&0&0&1&1&1&1&0\\
0&0&0&1&0&0&1&1&0&0\\
0&0&0&1&0&0&1&1&0&0\\

\end{bmatrix}
,\hspace{1cm}
E=\begin{bmatrix}

0&0&0&0&0&1&1&1&0&1\\
0&0&0&0&0&1&1&1&0&0\\
0&1&0&0&1&0&1&0&0&0\\
0&1&0&0&0&0&1&0&0&0\\
0&1&0&0&1&0&1&0&0&1\\
0&1&0&0&0&0&1&0&0&0\\
0&0&0&0&1&0&1&0&0&0\\
0&0&0&0&0&0&1&0&0&0\\
0&0&0&0&1&0&1&0&0&1\\
0&0&0&0&0&0&1&0&0&0\\

0&1&0&0&1&0&1&0&1&0\\
0&1&0&0&0&0&1&0&1&0\\
0&1&0&0&1&0&1&0&1&1\\
0&1&0&0&0&0&1&0&1&0\\
0&0&0&0&1&0&1&0&0&0\\
0&0&0&0&0&0&1&0&0&0\\
0&0&0&0&1&0&1&0&0&1\\
0&0&0&0&0&0&1&0&0&0\\
0&0&0&0&0&0&0&1&0&0\\
0&0&0&0&0&0&0&0&1&0\\
0&0&0&0&0&0&0&0&0&1
\end{bmatrix}.
\]

\vspace{8pt}
Then we can represent the constraint \eqref{constrain1} as $A\overrightarrow{q}=b$. Note that $\mathrm{HR}(T\rightarrow Y)=c'\overrightarrow{q}$. Thus the our problem can be equivalently rewritten as a minimization problem with linear constraints:
\begin{equation*}\label{LPmin}
\min~ c'\overrightarrow{q},~~\mathrm{s.t.}~~ \left\{
\begin{array}{ll}
A\overrightarrow{q}=b ,  \\
\overrightarrow{q}\geq 0 .
\end{array}
\right.
\end{equation*}  
and 
\begin{equation*}\label{LPmax}
\max~ c'\overrightarrow{q},~~\mathrm{s.t.}~~ \left\{
\begin{array}{ll}
A\overrightarrow{q}=b ,  \\
\overrightarrow{q}\geq0 .
\end{array}
\right.
\end{equation*}
In order to get the solution of the problem, we consider its dual problem:
\begin{equation}\label{LPmindual}
\max b'\overrightarrow{p},~~s.t.~~  A'\overrightarrow{p}\leq c .
\end{equation}
The set $Q=\{\overrightarrow{p}|A'\overrightarrow{p}\leq c\}$ is a polyhedron and the quality $b'\overrightarrow{p}$ reaches its extreme value at the vertexes of the polyhedron $Q$. Since both $A$ and $c$ do not involve any symbolics, we can enumerate all the vertexes $\{p_{1},\cdots,p_{K}\}$ of $Q$. Thus, the solution to the optimal problem \eqref{LPmindual} is $L_{\mathrm{HR}}=\max\{b'p_{1},\cdots,b'p_{K}\}.$
The bound can be attained since all the vertexes belong to the set $Q$.

\subsection*{The proof of $c_2$ does not affect the bound of $\mathrm{HR}(T\rightarrow Y)$}\label{c2bound}
In order to prove $c_2$ does not shink the bound of $\mathrm{HR}(T\rightarrow Y)$, we only need to prove that even setting $c_2=0$, the bound of $\mathrm{HR}(T\rightarrow Y)$ does not change as the bounds with smaller value of $c_2$ are nested within those with a larger value of $c_2$. From \eqref{constrain2}, we know $c_2=0$ is equivalent to:

\begin{equation}\label{constrain3}
\left\{
\begin{array}{ll}
$$\sum_{j=0,1,2,3} (q_{8,j}+q_{9,j}+q_{10,j}+q_{11,j})=0  ,  \\
$$\sum_{j=0,1,2,3} (q_{2,j}+q_{6,j}+q_{10,j}+q_{14,j})=0  . \\

\end{array}
\right.
\end{equation}
This indicates that all of $q_{i,j}$ in \eqref{constrain3} should be 0. Since $\mathrm{HR}(T\rightarrow Y)=q_{8,0}+q_{9,0}+q_{12,0}+q_{13,0}+q_{8,1}+q_{10,1}+q_{12,1}+q_{14,1}
+q_{4,2}+q_{5,2}+q_{12,2}+q_{13,2}+q_{4,3}+q_{6,3}+q_{12,3}+q_{14,3},$ $c_2$ can affect $\mathrm{HR}(T\rightarrow Y)$ through $q_{8,0}, q_{8,1}, q_{9,0}, q_{10,1}, q_{14,1}, q_{6,3}, q_{14,3}$.We show below that after setting all the $q_{i,j}=0$ in \eqref{constrain3}, we can still find a set of $\{\tilde q_{i,j}\}$, such that  $\mathrm{HR}(T\rightarrow Y)$ remains the same and satisfies \eqref{constrain1}.

Supposed that $\{q_{i,j}\}$ is a set of numbers satisfies \eqref{constrain1} and \eqref{constrain2}. It is easy to verify that after setting
\begin{equation*}\label{qtilde}
\left\{ 
\begin{array}{ll}
$$\tilde q_{12,0}=q_{12,0}+q_{8,0},\\
$$\tilde q_{13,0}=q_{13,0}+q_{9,0}, \\
$$\tilde q_{12,1}=q_{12,1}+q_{8,1}+q_{10,1}+q_{14,1}, \\
$$\tilde q_{12,3}=q_{12,3}+q_{6,3}+q_{14,3},\\
$$\tilde q_{15,0}=q_{10,0}+q_{11,0}+q_{14,0}+q_{15,0}, \\
$$\tilde q_{1,2}=q_{8,2}+q_{9,2}+q_{1,2},\\
$$\tilde q_{3,2}=q_{2,2}+q_{10,2}+q_{11,2}+q_{3,2},\\
\end{array}
\right.
\quad\quad
\left\{ 
\begin{array}{ll}
$$\tilde q_{15,2}=q_{6,2}+q_{14,2}+q_{15,2},\\
$$\tilde q_{0,1}=q_{2,1}+q_{6,1}+q_{0,1},\\
$$\tilde q_{10,1}=q_{9,1}+q_{11,1}+q_{10,1},\\
$$\tilde q_{3,0}=q_{2,0}+q_{6,0}+q_{3,0},\\
$$\tilde q_{0,3}=q_{2,3}+q_{8,3}+q_{10,3}+q_{0,3},\\
$$\tilde q_{1,3}=q_{9,3}+q_{11,3}+q_{1,3},\\
\end{array}
\right.
\end{equation*}

$\mathrm{HR}(T\rightarrow Y)$ does not change. Additionally, such parameterization  still satisfies \eqref{constrain1} and $\sum_{i=0}^{15} q_{i,2}+\alpha=c_{1}$, or equivalently, $P(Y,S|T)$ is unchanged and $\mathrm{HR}(T\rightarrow S)\leq c_1$ still holds. Hence for all $x\in [L,U]$, there exist a set of $\tilde q_{i,j}$  such that $\mathrm{HR}(T\rightarrow Y)=x$ and it satisfies $P(Y,S|T)$ is unchanged and $\mathrm{HR}(T\rightarrow S)\leq c_1$ still holds. Therefore, $c_2$ does not affect the bound of $\mathrm{HR(T\rightarrow Y)}$.

%\vspace{-4mm}
\newpage
\subsection*{Formula for $\tilde U_1,\ldots, \tilde U_{15}$}

Let $P_{ys|t}=P(Y=y,S=s|T=t)$, then, we have that
%\begin{landscape}
%	{$$\tilde U_{}=\min(\tilde U_1,\ldots,\tilde U_{15})^T=
%		$$}
	
\vspace{-4mm}
	{$$
		\tilde U_{}=\min(\tilde U_1,\ldots,\tilde U_{15})^T=\min\left(
		\begin{array}{c}
		P(Y=0|T=1)  \\
		P(Y=1|T=0)  \\
		c_1+P_{10|0}+P_{01|1}  \\
		1+c_1-P_{01|0}-P_{10|1}  \\
		c_1+c_3-P_{01|0}+P_{01|1} \\
		c_1+c_3+P_{10|0}-P_{10|1}  \\
		c_3+P_{11|0}+P_{01|1}  \\
		c_3+P_{10|0}+P_{00|1}  \\
		1+c_3-P_{01|0}-P_{11|1}  \\
		1+c_3-P_{00|0}-P_{10|1}\\
		0.5c_3+c_1+0.5P_{01|1}-0.5P_{11|1}-0.5P_{01|0}+0.5P_{11|0}  \\
		0.5+c_3-0.5P_{11|1}+0.5P_{01|1}+0.5P_{11|0}-0.5P_{01|0}  \\
		0.5+0.5c_3+0.5P_{00|1}-0.5P_{11|1}-0.5P_{01|0}+0.5P_{10|0}  \\
		0.5+0.5c_3+0.5P_{01|1}-0.5P_{10|1}+0.5P_{11|0}-0.5P_{00|0} \\
		0.5+0.5c_3+0.5P_{00|1}-0.5P_{10|1}-0.5P_{00|0}+0.5P_{10|0}  \\
		\end{array}
		\right).
		$$}
	
%\end{landscape}	

\vspace{-10mm}
\subsection*{The distribution of $P(Y|S,T)$ under linear model (A.1)}\label{chap3: pf1}
According to model \eqref{model_prentice_con}, we have
\begin{equation}
\left\{
\begin{array}{ll}
Y=\alpha_{0}+\alpha_{1}T+\alpha_{2}S+\alpha_{3}W+\epsilon_{1} ,  \\
S=\beta_{0}+\beta_{1}T+\beta_{2}W+\epsilon_{2} ,  \\
\epsilon_{1}\sim N(0,1),~~\epsilon_{2}\sim N(0,1),~~ W\sim N(0,1),~~ (\epsilon_{1},\epsilon_{2})\bot (T,S,W),~~ \epsilon_{1}\bot\epsilon_{2} .
\end{array}
\right.
\end{equation}
Then, we can calculate the density of $W$ given $(S,T)$ as
\begin{eqnarray*}
	P(W=w|S=s,T=t) &\propto& P(S=s|T=t,W=w)P(W=w) \\
	&\propto& \exp\Big\{-\frac{(s-\beta_{0}-\beta_{1}t-\beta_{2}w)^{2}}{2}\Big\}\exp\Big\{-\frac{w^{2}}{2}\Big\} \\
	&\propto& \exp\Big\{-\frac{(1+\beta_{2}^{2})w^{2}-2(s-\beta_{0}-\beta_{1}t)\beta_{2}w}{2}\Big\} \\
	&\propto& \exp\Big\{\frac{(1+\beta_{2}^{2})\big(w-\frac{\beta_{2}(s-\beta_{0}-\beta_{1}t)}{1+\beta_{2}^{2}}\big)^{2}}{2}\Big\}.
\end{eqnarray*}
So we have $W|(S,T)\sim N\{\beta_{2}(S-\beta_{0}-\beta_{1}T)/(1+\beta_{2}^{2}), 1/(1+\beta_{2}^{2})\}.$
%\newline

\vspace{-5mm}
\begin{eqnarray*}
	&&P(Y=y|S=s,T=t)\\ &\propto& \int P(Y=y|S=s, T=t, W=w)P(W=w|S=s,T=t) dw \\
	&\propto& \int \exp\Big\{-\frac{(y-\alpha_{0}-\alpha_{1}t-\alpha_{2}s-\alpha_{3}w)^{2}}{2}\Big\}
	\exp\Big\{(1+\beta_{2}^{2})\big(w-\frac{\beta_{2}(s-\beta_{0}-\beta_{1}t)}{1+\beta_{2}^{2}}\big)^{2}\Big\}dw \\
	&=&\exp\Big\{-\frac{(y-\alpha_{0}-\alpha_{1}t-\alpha_{2}s)^{2}
		+(s-\beta_{0}-\beta_{1}t)^{2}\beta_{2}^{2}/(1+\beta_{2}^{2})}{2}\Big\}\cdot \\
	&&\int\exp\Big\{-\frac{(1+\beta_{2}^{2}+\alpha_{3}^{2})w^{2}
		-2\big[\alpha_{3}(y-\alpha_{0}-\alpha_{1}t-\alpha_{2}s)+\beta_{2}(s-\beta_{0}-\beta_{1}t)\big]w}{2}\Big\}dw\\
	&=&\exp\Big\{-\frac{(y-\alpha_{0}-\alpha_{1}t-\alpha_{2}s)^{2}
		+(s-\beta_{0}-\beta_{1}t)^{2}\beta_{2}^{2}/(1+\beta_{2}^{2})}{2}\Big\}\cdot \\
	&&\int\exp\Big\{-\frac{(1+\beta_{2}^{2}+\alpha_{3}^{2})
		\big(w-\big[\alpha_{3}(y-\alpha_{0}-\alpha_{1}t-\alpha_{2}s)+\beta_{2}(s-\beta_{0}-\beta_{1}t)\big]\big)^{2}}{2(1+\beta_{2}^{2}+\alpha_{3}^{2})}\Big\}dw\cdot\\
	&&\exp\Big\{-\frac{-\big[\alpha_{3}(y-\alpha_{0}-\alpha_{1}t-\alpha_{2}s)+\beta_{2}(s-\beta_{0}-\beta_{1}t)\big]^{2}}{2(1+\beta_{2}^{2}+\alpha_{3}^{2})}\Big\}\\
	&\propto&\exp\Big\{-\frac{(y-\alpha_{0}-\alpha_{1}t-\alpha_{2}s)^{2}
		-\big[\alpha_{3}(y-\alpha_{0}-\alpha_{1}t-\alpha_{2}s)+\beta_{2}(s-\beta_{0}-\beta_{1}t)\big]^{2}}{2(1+\beta_{2}^{2}+\alpha_{3}^{2})}\Big\}\\
	&\propto&\exp\Big\{-\frac{(1+\beta_{2}^{2})y^{2}
		-2\big(\alpha_{3}\beta_{2}(s-\beta_{0}-\beta_{1}t)+(1+\beta_{2}^{2})(\alpha_{0}+\alpha_{1}t+\alpha_{2}s)\big)y}{2(1+\beta_{2}^{2}+\alpha_{3}^{2})}\Big\}\\
	&\propto&\exp\Big\{-\frac{(1+\beta_{2}^{2})
		\big(y-\mu(t,s)\big)^{2}}{2(1+\beta_{2}^{2}+\alpha_{3}^{2})}\Big\},
\end{eqnarray*}

\noindent where $\mu(t,s)=\big[\big\{(1+\beta_{2}^{2})\alpha_{0}-\alpha_{3}\beta_{0}\beta_{2}\big\}
+\big\{\alpha_{1}(1+\beta_{2}^{2})-\alpha_{3}\beta_{1}\beta_{2}\big\}t
+\big\{\alpha_{2}(1+\beta_{2}^{2})+\alpha_{3}\beta_{2}\big)s\big]/(1+\beta_{2}^{2}\}.$ Hence, we have
$$Y|(T,S)\sim N\{\mu(T,S), ~\frac{1+\beta_{2}^{2}}{2(1+\beta_{2}^{2}+\alpha_{3}^{2})}\}.$$

\subsection*{Tables in the paper}
\begin{table}[!ht]
	\caption{{The 64 categories of the subgroups when $S$ and $Y$ are binary and $S$ is a non-strong surrogate.}\label{tb: non_strong_q}}
	\begin{center}
		\begin{tabular}{|l|l|l|l|l|}
			\hline
			& $S_{0}=0,S_{1}=0$ & $S_{0}=0,S_{1}=1$ & $S_{0}=1,S_{1}=0$ & $S_{0}=1,S_{1}=1$ \\
			\hline
			$(Y_{00},Y_{01},Y_{10},Y_{11})=(0,0,0,0)$ & $q_{0,0}$ & $q_{0,1}$ & $q_{0,2}$ & $q_{0,3}$ \\
			\hline
			$(Y_{00},Y_{01},Y_{10},Y_{11})=(0,0,0,1)$ & $q_{1,0}$ & $q_{1,1}$ & $q_{1,2}$ & $q_{1,3}$ \\
			\hline
			$(Y_{00},Y_{01},Y_{10},Y_{11})=(0,0,1,0)$ & $q_{2,0}$ & $q_{2,1}$ & $q_{2,2}$ & $q_{2,3}$ \\
			\hline
			$(Y_{00},Y_{01},Y_{10},Y_{11})=(0,0,1,1)$ & $q_{3,0}$ & $q_{3,1}$ & $q_{3,2}$ & $q_{3,3}$ \\
			\hline
			$(Y_{00},Y_{01},Y_{10},Y_{11})=(0,1,0,0)$ & $q_{4,0}$ & $q_{4,1}$ & $q_{4,2}$ & $q_{4,3}$ \\
			\hline
			$(Y_{00},Y_{01},Y_{10},Y_{11})=(0,1,0,1)$ & $q_{5,0}$ & $q_{5,1}$ & $q_{5,2}$ & $q_{5,3}$ \\
			\hline
			$(Y_{00},Y_{01},Y_{10},Y_{11})=(0,1,1,0)$ & $q_{6,0}$ & $q_{6,1}$ & $q_{6,2}$ & $q_{6,3}$ \\
			\hline
			$(Y_{00},Y_{01},Y_{10},Y_{11})=(0,1,1,1)$ & $q_{7,0}$ & $q_{7,1}$ & $q_{7,2}$ & $q_{7,3}$ \\
			\hline
			$(Y_{00},Y_{01},Y_{10},Y_{11})=(1,0,0,0)$ & $q_{8,0}$ & $q_{8,1}$ & $q_{8,2}$ & $q_{8,3}$ \\
			\hline
			$(Y_{00},Y_{01},Y_{10},Y_{11})=(1,0,0,1)$ & $q_{9,0}$ & $q_{9,1}$ & $q_{9,2}$ & $q_{9,3}$ \\
			\hline
			$(Y_{00},Y_{01},Y_{10},Y_{11})=(1,0,1,0)$ & $q_{10,0}$ & $q_{10,1}$ & $q_{10,2}$ & $q_{10,3}$ \\
			\hline
			$(Y_{00},Y_{01},Y_{10},Y_{11})=(1,0,1,1)$ & $q_{11,0}$ & $q_{11,1}$ & $q_{11,2}$ & $q_{11,3}$ \\
			\hline
			$(Y_{00},Y_{01},Y_{10},Y_{11})=(1,1,0,0)$ & $q_{12,0}$ & $q_{12,1}$ & $q_{12,2}$ & $q_{12,3}$ \\
			\hline
			$(Y_{00},Y_{01},Y_{10},Y_{11})=(1,1,0,1)$ & $q_{13,0}$ & $q_{13,1}$ & $q_{13,2}$ & $q_{13,3}$ \\
			\hline
			$(Y_{00},Y_{01},Y_{10},Y_{11})=(1,1,1,0)$ & $q_{14,0}$ & $q_{14,1}$ & $q_{14,2}$ & $q_{14,3}$ \\
			\hline
			$(Y_{00},Y_{01},Y_{10},Y_{11})=(1,1,1,1)$ & $q_{15,0}$ & $q_{15,1}$ & $q_{15,2}$ & $q_{15,3}$ \\
			\hline
		\end{tabular}
	\end{center}
\end{table}

\begin{table}[!htb]
	\caption{The 27 categories of the subgroups when surrogate $S$ and outcome $Y$ are binary, $T$ is beneficial to $S$ and $S$ is beneficial to $Y$.}\label{tb: nonstrong_q_noindi_para}
	\begin{center}
		\begin{tabular}{|l|l|l|l|}
			\hline
			& $S_{0}=0,S_{1}=0$ & $S_{0}=0,S_{1}=1$  & $S_{0}=1,S_{1}=1$ \\
			\hline
			$(Y_{00},Y_{01},Y_{10},Y_{11})=(0,0,0,0)$ & $q_{0,0}$ & $q_{0,1}$ &  $q_{0,3}$ \\
			\hline
			$(Y_{00},Y_{01},Y_{10},Y_{11})=(0,0,0,1)$ & $q_{1,0}$ & $q_{1,1}$ &  $q_{1,3}$ \\
			\hline
			$(Y_{00},Y_{01},Y_{10},Y_{11})=(0,0,1,1)$ & $q_{3,0}$ & $q_{3,1}$ &  $q_{3,3}$ \\
			\hline
			$(Y_{00},Y_{01},Y_{10},Y_{11})=(0,1,0,0)$ & $q_{4,0}$ & $q_{4,1}$ &  $q_{4,3}$ \\
			\hline
			$(Y_{00},Y_{01},Y_{10},Y_{11})=(0,1,0,1)$ & $q_{5,0}$ & $q_{5,1}$ &  $q_{5,3}$ \\
			\hline
			$(Y_{00},Y_{01},Y_{10},Y_{11})=(0,1,1,1)$ & $q_{7,0}$ & $q_{7,1}$ &  $q_{7,3}$ \\
			\hline
			$(Y_{00},Y_{01},Y_{10},Y_{11})=(1,1,0,0)$ & $q_{12,0}$ & $q_{12,1}$ &  $q_{12,3}$ \\
			\hline
			$(Y_{00},Y_{01},Y_{10},Y_{11})=(1,1,0,1)$ & $q_{13,0}$ & $q_{13,1}$ &  $q_{13,3}$ \\
			\hline
			$(Y_{00},Y_{01},Y_{10},Y_{11})=(1,1,1,1)$ & $q_{15,0}$ & $q_{15,1}$ &  $q_{15,3}$ \\
			\hline
		\end{tabular}
	\end{center}
\end{table}

\begin{table}[!htb]
	\caption{The counterexample of Prentice criterion when surrogate $S$ and outcome $Y$ are binary.}\label{tb: eg_prentice}
	\begin{center}
		\begin{tabular}{|l|l|l|l|}
			\hline
			& $S_{0}=0,S_{1}=0$ & $S_{0}=0,S_{1}=1$  & $S_{0}=1,S_{1}=1$ \\
			\hline
			$(Y_{00},Y_{01},Y_{10},Y_{11})=(0,0,0,0)$ & 0.01 & 0.015 &  0.055 \\
			\hline
			$(Y_{00},Y_{01},Y_{10},Y_{11})=(0,0,0,1)$ & 0.02 & 0.02 &  0.025 \\
			\hline
			$(Y_{00},Y_{01},Y_{10},Y_{11})=(0,0,1,1)$ & 0.04 & 0.05 &  0.07 \\
			\hline
			$(Y_{00},Y_{01},Y_{10},Y_{11})=(0,1,0,0)$ & 0 & 0.015 &  0.05 \\
			\hline
			$(Y_{00},Y_{01},Y_{10},Y_{11})=(0,1,0,1)$ & 0 & 0.04 &  0.03 \\
			\hline
			$(Y_{00},Y_{01},Y_{10},Y_{11})=(0,1,1,1)$ & 0.03 & 0.06 &  0.065 \\
			\hline
			$(Y_{00},Y_{01},Y_{10},Y_{11})=(1,1,0,0)$ & 0.05 & 0.08 &  0.035 \\
			\hline
			$(Y_{00},Y_{01},Y_{10},Y_{11})=(1,1,0,1)$ & 0.02 & 0.07 &  0.045 \\
			\hline
			$(Y_{00},Y_{01},Y_{10},Y_{11})=(1,1,1,1)$ & 0.03 & 0.05 &  0.075 \\
			\hline
		\end{tabular}
	\end{center}
\end{table}

\begin{table}[!htb]
	\caption{The counterexample of principle surrogate when surrogate $S$ and outcome $Y$ are binary.}\label{counter_principal_bin}
	\begin{center}
		\begin{tabular}{|l|l|l|l|}
			\hline
			& $S_{0}=0,S_{1}=0$ & $S_{0}=0,S_{1}=1$  & $S_{0}=1,S_{1}=1$ \\
			\hline
			$(Y_{00},Y_{01},Y_{10},Y_{11})=(0,0,0,0)$ & 0.025 & 0.015 &  0.005 \\
			\hline
			$(Y_{00},Y_{01},Y_{10},Y_{11})=(0,0,0,1)$ & 0.03 & 0.02 &  0.025 \\
			\hline
			$(Y_{00},Y_{01},Y_{10},Y_{11})=(0,0,1,1)$ & 0.04 & 0.05 &  0.06 \\
			\hline
			$(Y_{00},Y_{01},Y_{10},Y_{11})=(0,1,0,0)$ & 0.035 & 0.015 &  0.05 \\
			\hline
			$(Y_{00},Y_{01},Y_{10},Y_{11})=(0,1,0,1)$ & 0.04 & 0.04 &  0.03 \\
			\hline
			$(Y_{00},Y_{01},Y_{10},Y_{11})=(0,1,1,1)$ & 0.03 & 0.06 &  0.015 \\
			\hline
			$(Y_{00},Y_{01},Y_{10},Y_{11})=(1,1,0,0)$ & 0.05 & 0.08 &  0.035 \\
			\hline
			$(Y_{00},Y_{01},Y_{10},Y_{11})=(1,1,0,1)$ & 0.02 & 0.07 &  0.025 \\
			\hline
			$(Y_{00},Y_{01},Y_{10},Y_{11})=(1,1,1,1)$ & 0.03 & 0.05 &  0.055 \\
			\hline
		\end{tabular}
	\end{center}
\end{table}

\begin{table}[!ht]
	\caption{{The table used by counterexample in Example \ref{ex:strong}.}\label{tb: counterexample_q}}
	\begin{center}
		\begin{tabular}{|l|l|l|l|l|l|l|l|l|}
			\hline
			&&&\multicolumn{3}{|c|}{$(Y_0,Y_1,Y_2)$}&&&\\
			& $(0,0,0)$ & $(0,0,1)$ & $(0,1,0)$ & $(0,1,1)$ & $(1,0,0)$ & $(1,0,1)$ & $(1,1,0)$ & $(1,1,1)$\\
			\hline
			$(S_0=0, S_1=0)$ & $q_{0,0}$ & $q_{0,1}$ & $q_{0,2}$ & $q_{0,3}$ & $q_{0,4}$ & $q_{0,5}$ & $q_{0,6}$ & $q_{0,7}$\\
			\hline
			$(S_0=0, S_1=1)$ & $q_{1,0}$ & $q_{1,1}$ & $q_{1,2}$ & $q_{1,3}$ & $q_{1,4}$ & $q_{1,5}$ & $q_{1,6}$ & $q_{1,7}$\\
			\hline
			$(S_0=0, S_1=2)$ & $q_{2,0}$ & $q_{2,1}$ & $q_{2,2}$ & $q_{2,3}$ & $q_{2,4}$ & $q_{2,5}$ & $q_{2,6}$ & $q_{2,7}$\\
			\hline
			$(S_0=1, S_1=0)$ & $q_{3,0}$ & $q_{3,1}$ & $q_{3,2}$ & $q_{3,3}$ & $q_{3,4}$ & $q_{3,5}$ & $q_{3,6}$ & $q_{3,7}$\\
			\hline
			$(S_0=1, S_1=1)$ & $q_{4,0}$ & $q_{4,1}$ & $q_{4,2}$ & $q_{4,3}$ & $q_{4,4}$ & $q_{4,5}$ & $q_{4,6}$ & $q_{4,7}$\\
			\hline
			$(S_0=1, S_1=2)$ & $q_{5,0}$ & $q_{5,1}$ & $q_{5,2}$ & $q_{5,3}$ & $q_{5,4}$ & $q_{5,5}$ & $q_{5,6}$ & $q_{5,7}$\\
			\hline
			$(S_0=2, S_1=0)$ & $q_{6,0}$ & $q_{6,1}$ & $q_{6,2}$ & $q_{6,3}$ & $q_{6,4}$ & $q_{6,5}$ & $q_{6,6}$ & $q_{6,7}$\\
			\hline
			$(S_0=2, S_1=1)$ & $q_{7,0}$ & $q_{7,1}$ & $q_{7,2}$ & $q_{7,3}$ & $q_{7,4}$ & $q_{7,5}$ & $q_{7,6}$ & $q_{7,7}$\\
			\hline
			$(S_0=2, S_1=2)$ & $q_{8,0}$ & $q_{8,1}$ & $q_{8,2}$ & $q_{8,3}$ & $q_{8,4}$ & $q_{8,5}$ & $q_{8,6}$ & $q_{8,7}$\\
			\hline
			
		\end{tabular}
	\end{center}
\end{table}

\label{lastpage}

\end{document}